\documentclass[a4paper,11pt]{article}
\pdfoutput=1
\usepackage{jheppub}
\usepackage[T1]{fontenc}
\usepackage[utf8]{inputenc}

\usepackage{braket}
\usepackage{bbold}
\usepackage{subcaption}

\usepackage{amsthm}
\usepackage{amsmath}

\newtheorem{theorem}{Theorem}[section]
\newtheorem{lemma}[theorem]{Lemma}

\graphicspath{ {./plots/} }

\title{\boldmath Scalar conformal primary fields in the Brownian loop soup}

\author[a,b,c]{Federico Camia}
\author[a]{Valentino F.\ Foit}
\author[a]{Alberto Gandolfi}
\author[d]{Matthew Kleban}

\affiliation[a]{Science Division, New York University Abu Dhabi, Saadiyat Island, Abu Dhabi, United Arab Emirates}
\affiliation[b]{Courant Institute of Mathematical Sciences, New York University, 251 Mercer Street, New York, NY 10012, USA}
\affiliation[c]{Department of Mathematics, Vrije Universiteit Amsterdam, De Boelelaan 1111, 1081 HV Amsterdam, The Netherlands}
\affiliation[d]{Center for Cosmology and Particle Physics, New York University, 726 Broadway, New York, NY 10003, USA}

\emailAdd{federico.camia@nyu.edu}
\emailAdd{foit@nyu.edu}
\emailAdd{albertogandolfi@nyu.edu}
\emailAdd{kleban@nyu.edu}

\abstract{The Brownian loop soup is a conformally invariant statistical ensemble of random loops in two dimensions characterized by an intensity $\lambda>0$, with central charge $c=2 \lambda$. Recent progress resulted in an analytic form for the four-point function of a class of scalar conformal primary ``layering vertex operators'' $\mathcal{O}_{\beta}$ with  dimensions $(\Delta, \Delta)$, with $\Delta = \frac{\lambda}{10}(1-\cos\beta)$, that compute certain statistical properties of the model.
The Virasoro conformal block expansion of the four-point function revealed the existence of a new set of operators with dimensions $(\Delta+ k/3, \Delta + k'/3)$, for all non-negative integers $k, k'$ satisfying $|k-k'| = 0$ mod 3.
In this paper we introduce the edge counting field $\mathcal E(z)$ that counts the number of loop boundaries that pass close to the point $z$. We rigorously prove that the $n$-point functions of $\mathcal E$ are well defined and behave as expected for a conformal primary field with dimensions $(1/3, 1/3)$. We analytically compute the four-point function $\Braket{\mathcal{O}_{\beta}(z_1) \mathcal{O}_{-\beta}(z_2) \mathcal{E}(z_3) \mathcal{E}(z_4)}$ and analyze its conformal block expansion. The operator product expansions of $\mathcal{E} \times \mathcal{E}$ and $\mathcal{E} \times \mathcal{O}_{\beta}$  produce higher-order edge operators with ``charge'' $\beta$ and dimensions $(\Delta + k/3, \Delta + k/3)$. Hence, we have explicitly identified all scalar primary operators among the new set  mentioned above.
We also re-compute the central charge by an independent method based on the operator product expansion and find agreement with previous methods.
}

\begin{document}
\maketitle
\flushbottom

\section{Introduction}

The Brownian loop soup (BLS) \cite{lawler2004brownian} is an ideal gas of Brownian loops with a distribution chosen so that it is invariant under local conformal transformations. The BLS is implicit in the work of Symanzik \cite{osti_4117149} on Euclidean quantum field theory, more precisely, in the representation of correlation functions of Euclidean fields in terms of random paths that are locally statistically equivalent to Brownian motion. This representation can be made precise for the Gaussian free field, in which case the random paths are independent of each other and can be generated as a Poisson process.

The BLS is closely related not only to Brownian motion and the Gaussian free field but also to the Schramm-Loewner Evolution (SLE) and Conformal Loop Ensembles (CLEs). It provides an interesting and useful link between Brownian motion, field theory, and statistical mechanics. Partly motivated by these connections, as well as by a potential application to cosmology in the form of a conformal field theory for eternal inflation \cite{Freivogel:2009rf}, three of the present authors introduced a set of operators that compute properties of the BLS and discovered new families of conformal primary fields depending on a real parameter $\beta$ \cite{Camia_2016}. One such family are the fields $\mathcal{O}_{\beta}$.  These operators have scaling dimensions $\Delta(\beta) = \frac{\lambda}{10}(1-\cos\beta)$ and are periodic under $\beta \rightarrow \beta + 2 \pi$, with $\mathcal{O}_{0} \equiv \mathcal{O}_{2 \pi} = \mathbb{1}$ (the identity operator).
Their $n$-point function $\Braket{\mathcal{O}_{\beta_1}(z_1) \ldots \mathcal{O}_{\beta_n}(z_n)}_{\mathbb{C}}$ in the full plane is identically zero unless $\sum_{j=1}^n \beta_j = 0 \; \mod{2\pi}$, which is reminiscent of the ``charge neutrality'' or ``charge conservation'' condition that applies to vertex operators of the free boson \cite{DiFrancesco:639405}.

These operators were further studied in \cite{Camia_2020}, where it is shown that the operator product expansion (OPE) $\mathcal{O}_{\beta_i} \times \mathcal{O}_{\beta_j}$ predicts the existence of operators of dimensions $(\Delta_{ij} + \frac{k}{3}, \Delta_{ij} + \frac{k'}{3})$ for all non-negative integers $k, k'$  satisfying $|k-k'| = 0$ mod 3, where $\Delta_{ij} = \frac{\lambda}{10}(1-\cos(\beta_i + \beta_j))$.  The simplest case is $k=k'=1$ and $\beta_i + \beta_j = 0$ mod $2 \pi$ so that $\Delta_{ij}=0$ and the dimensions are $(1/3,1/3)$. These results were derived by exploiting a connection between the BLS and the $O(n)$ model in the limit $n \to 0$. Further generalizations of the layering operators were explored in \cite{foit2020new}.

While the analysis in \cite{Camia_2020} demonstrated that  new operators must exist and allowed us to compute their dimensions and three-point function coefficients with $\mathcal{O}_{\beta}$, it did not provide a clue as to how they are defined in terms of loops of the BLS loop ensemble.  In this paper we introduce a new field $\mathcal E(z)$ that counts the number of outer boundaries of BLS loops that pass close to $z$ and rigorously prove that its $n$-point functions are well defined and behave as expected for a primary field. We  identify $\mathcal E$ with the operator of dimensions $(1/3,1/3)$ discovered in \cite{Camia_2020}, compute the four-point function $\Braket{\mathcal{O}_{\beta}(z_1) \mathcal{O}_{-\beta}(z_2) \mathcal{E}(z_3) \mathcal{E}(z_4)}_\mathbb{C}$, and perform its Virasoro conformal block expansion.  This provides further information about three-point function coefficients and the spectrum of primary operators. We further define higher order ($k=k'>1$) and charged ($\beta \neq 0$) generalizations of this operator that can be  identified with the operators of dimensions $(\Delta_{ij} + \frac{k}{3}, \Delta_{ij} + \frac{k}{3})$. In other words, we identify and explicitly define in terms of the loops all spin-zero primary fields emerging from the Virasoro conformal block expansion derived in \cite{Camia_2020}. 

This corpus of results establishes the BLS as a novel conformal field theory (CFT), or class of conformal field theories, with certain unique features (such as the periodicity of the operator dimensions in the charge $\beta$). Nevertheless, many aspects of this CFT remain mysterious---among other things, the nature of the operators with non-zero spin, $|k - k'| \neq 0$.  The relation of this CFT to other better-known CFTs and its possible role as a model for physical phenomena also  remains unclear.

\subsection{Preliminary definitions} \label{sec:definitions}

If $A$ is a set of loops in a domain $D$, the partition function of the BLS restricted to loops from $A$ can be written as
\begin{equation}
    {Z_A = \sum_{n=0}^{\infty} \frac{\lambda^n}{n!} \left( \mu_D^{\text{loop}}(A) \right)^n},
\end{equation}
where $\lambda>0$ is a constant and $\mu_D^{\text{loop}}$ is a measure on planar loops in $D$ called \emph{Brownian loop measure} and defined as
\begin{equation} \label{brownian-loop-measure}
\mu_D^{\text{loop}} := \int_{D} \int_0^{\infty} \frac{1}{2 \pi t^2} \, \mu^{br}_{z,t} \, dt \, d{\bf A}(z),
\end{equation}
where $\bf A$ denotes area and $\mu^{br}_{z,t}$ is the complex Brownian bridge measure with starting point
$z$ and duration $t$.\footnote{We note that the Brownian loop measure should be interpreted as a measure on ``unrooted''
loops, that is, loops without a specified starting point. Unrooted loops are equivalence classes of rooted
loops. The interested reader is referred to \cite{lawler2004brownian} for more details.}
$Z_A$ can be thought of as the grand canonical partition function of a system of loops with fugacity $\lambda$, and the BLS can be shown to be conformally invariant and to have central charge $c=2\lambda$ (see \cite{lawler2004brownian,Camia_2016}).  

In this paper we will only be concerned with the outer boundaries of Brownian loops. More precisely, given a planar loop $\gamma$ in $\mathbb C$, its outer boundary or ``edge'' $\ell=\ell(\gamma)$ is the boundary of the unique infinite component of ${\mathbb C} \setminus \gamma$. Note that, for any planar loop $\gamma$, $\ell(\gamma)$ is always a simple closed curve, i.e., a closed loop without self-intersections. Hence, in this paper, we will work with collections ${\mathcal L}$ of simple loops $\ell$ which are the outer boundaries of the loops from a BLS and for us, with a slight abuse of terminology, a BLS will be a collection of simple loops.  With these understandings, the $\lambda \to 0$ limit (interpreted appropriately) reduces to the case of a single self-avoiding loop.  There is a unique (up to an overall multiplicative constant) conformally invariant measure on such loops \cite{2005math.....11605W}, which are also described by the $n \to 0$ limit of the $O(n)$ model. Exploiting this connection allowed us to obtain exact results for certain correlation functions here and in our previous work \cite{Camia_2020}.

Given a simple loop $\ell$, let $\bar\ell$ denote its \emph{interior}, i.e.\ the unique bounded simply connected component of $\mathbb{C} \setminus \ell$. In other words, a point $z$ belongs to $\bar\ell$ if $\ell$ disconnects $z$ from infinity, in which case we write $z \in \bar\ell$. In \cite{Camia_2016}, the authors studied the correlation functions of the \emph{layering operator} or \emph{field}\footnote{In this paper we  use the terms \emph{field} and \emph{operator} interchangeably.} $V_{\beta}(z)=\exp({i\beta\sum_{\ell: z \in \bar\ell} \sigma_{\ell}}(z))$, where $\sigma_{\ell}$ are independent, symmetric, $(\pm 1)$-valued Boolean variables associated to the loops. One difficulty arises immediately due to the scale invariance of the BLS, which implies that the sum at the exponent is infinite with probability one. This difficulty can be overcome by imposing a short-distance cutoff $\delta>0$ 
on the diameter of loops (essentially removing from the loop soup all loops with diameter smaller than $\delta$.\footnote{An additional infrared cutoff or a ``charge neutrality'' or ``charge conservation'' condition may be necessary in some circumstances---we refer the interested reader to \cite{Camia_2016} for more details.}) As shown in \cite{Camia_2016}, the cutoff $\delta$ can be removed by rescaling the cutoff version $V^{\delta}_{\beta}$ of $V_{\beta}$ by $\delta^{-2\Delta(\beta)}$ and sending $\delta \to 0$. When $\delta \to 0$, the $n$-point correlation functions of $\delta^{-2\Delta(\beta)}V^{\delta}_{\beta}$ converge to conformally covariant quantities \cite{Camia_2016}, showing that the limiting field is a scalar conformal primary field with real and positive scaling dimension varying continuously as a periodic function of $\beta$, namely as $\Delta(\beta) = \bar \Delta(\beta) = \frac{\lambda}{10}(1-\cos\beta)$. This limiting field is further studied in \cite{Camia_2020}, where its canonically normalized version is denoted by $\mathcal{O}_{\beta}$.\footnote{By \emph{canonically normalized} we mean that the full-plane two-point function $\Braket{\mathcal{O}_{\beta}(z) \mathcal{O}_{-\beta}(z')}_{\mathbb C} = |z-z'|^{-2\Delta(\beta)}$.}

\begin{figure}[t]
    \centering
    \includegraphics[width=.75\textwidth]{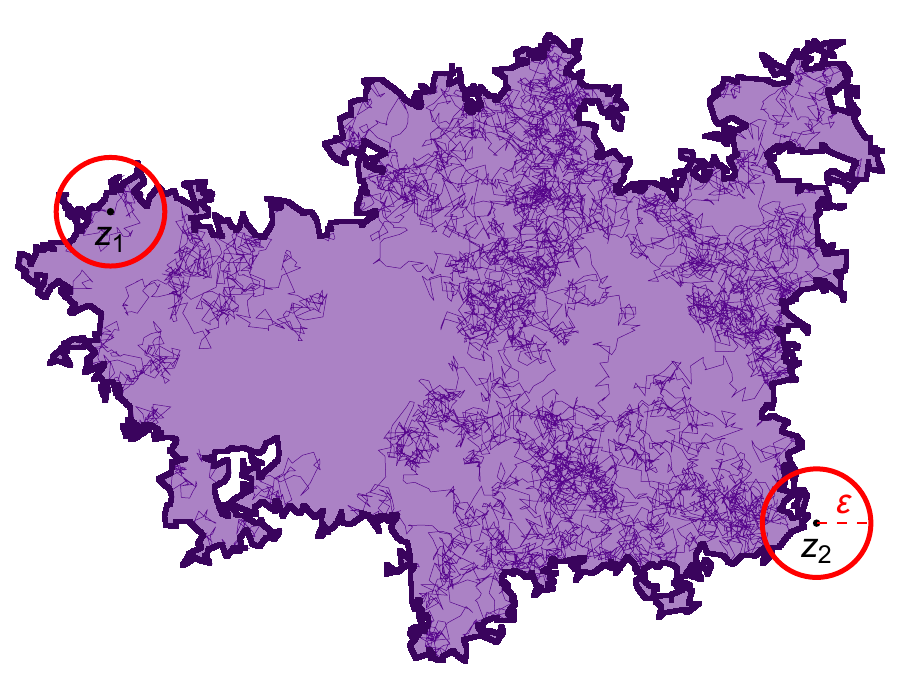}
    \caption{A Brownian loop (thin NYU violet line) and its boundary (thick violet line; the interior is shaded). Such a loop would contribute to the two-point function of edge operators inserted at $z_{1}$ and $z_2$ because the loop comes within $\varepsilon$ of both. It would contribute to a layering operator inserted at $z_1$ (but not $z_2$) because $z_1$ (but not $z_2$) is in the interior of the loop (that is, the loop separates $z_1$ from infinity, but not $z_2$).}
    \label{edge_edge}
\end{figure}

The \emph{edge field} $\mathcal{E}(z)$ studied in this paper counts the number of loops $\ell$ passing within a short-distance  $\varepsilon$  of the point $z$. 
The cutoff and renormalization procedure described in Section \ref{sec:edge}  shows that $\mathcal{E}$ has well defined $n$-point functions which are conformally covariant, and that it behaves like a scalar conformal primary with  scaling dimension $(1/3,1/3)$.
This scaling dimension can be understood qualitatively as follows.  It is known that the fractal dimension of the boundary of a Brownian loop is 4/3 \cite{2000math.....10165L}.  Fattening the loop's boundary into a  strip\footnote{Recipes for Wiener sausages in Brownian  soups are available on special request.} of width $\varepsilon$, a fractal dimension of 4/3 means that the area of the strip is proportional to $\varepsilon^{2/3}$. Hence the probability for a loop to come within $\varepsilon$ of a given point scales as $\varepsilon^{2/3}$.  Loops that contribute to the two-point function of the edge operator with itself must come close to both points (Figure \ref{edge_edge}).  Therefore the two-point function is proportional to the square of this probability $\left| \varepsilon/z_{12} \right|^{4/3}$, where the power of $z_{12}$ follows from invariance under an overall scale transformation $(\varepsilon,z) \rightarrow (\lambda \varepsilon, \lambda z)$.  This dependence on $|z_{12}|$ is that of a scalar operator with dimension $(1/3,1/3)$.

In Section \ref{sec:higher_order} we identify additional scalar fields resulting from combinations of the edge field $\mathcal{E}$ with itself that we denote by $\mathcal{E}^{(k)}$ and call \emph{higher-order} edge operators. These fields have holomorphic and anti-holomorphic dimension $\frac{k}{3}$ for all non-negative integers $k$. In Section \ref{sec:charged} we discuss ``charged'' versions of the (higher-order) edge operators resulting from combinations of the edge field with itself and with the layering field $\mathcal{O}_{\beta}$; we denote these by $\mathcal{E}^{(k)}_{\beta}$ and call them \emph{charged} edge operators. These fields have holomorphic and anti-holomorphic dimension $\Delta(\beta) + \frac{k}{3}$, with non-negative integer $k$. The higher-order and charged edge operators complete the list of all scalar primary fields in  the conformal block expansion derived in \cite{Camia_2020}.

\subsection{Summary of the main results}

The domains $D$ considered in this paper are the full (complex) plane $\mathbb C$, the upper-half plane $\mathbb H$ or any domain conformally equivalent to $\mathbb H$. In this section and in the rest of the paper, we use $\Braket{\cdot}_D$ to denote expectation with respect to the BLS in $D$. The domain will be explicitly present in our notation when we want to emphasize its role; if the domain is not denoted in a particular expression (for example, if we use $\Braket{\cdot}$ instead of $\Braket{\cdot}_D$ or $\mu^{\text{loop}}$ instead of $\mu_D^{\text{loop}}$), it means that that expression is valid for any of the domains mentioned above.

The first group of main results concerns the Brownian loop measure $\mu^{\text{loop}}_D$ in a domain $D$, the $n$-point functions of the edge operator $\mathcal E$, which can be expressed in terms of $\mu^{\text{loop}}_D$, and the relation between $\mathcal E$ and $\mathcal{O}_{\beta}$.\footnote{The edge operator is properly defined in Section \ref{sec:edge} below.}
To formulate the results, we let $\vartheta_{\varepsilon}$ denote the scaling limit of the probability that, in critical site percolation on the triangular lattice, there are one open and two closed paths crossing the annulus with inner radius $\varepsilon$ and outer radius $1$, known as a \emph{three-arm event}. The existence of the limit is guaranteed by the existence of the full scaling limit of critical percolation \cite{Camia_2006}, and it is known that
$\vartheta_{\varepsilon} \sim \varepsilon^{2/3}$ (see Lemma \ref{lemma:theta} for a precise statement).

\begin{itemize}
    \item For any collection of distinct points $z_1,\ldots,z_k \in D$ with $k \geq 2$, letting $B_{\varepsilon}(z_j)$ denote the disk of radius $\varepsilon$ centered at $z_j$, the following limit exists
\begin{equation}
    \alpha_D^{z_1,\ldots,z_k} := \lim_{\varepsilon \to 0} \vartheta_{\varepsilon}^{-k} \mu^{\text{loop}}_D(\ell \cap B_{\varepsilon}(z_j) \neq \emptyset \; \forall j=1,\ldots,k).
\end{equation}
Moreover, $\alpha_D^{z_1,\ldots,z_k}$ is conformally covariant in the sense that, if $D'$ is a domain conformally equivalent to $D$ and $f:D \to D'$ is a conformal map, then
\begin{equation}
    \alpha_{D'}^{f(z_1),\ldots,f(z_k)} = \left( \prod_{j=1}^k |f'(z_j)|^{-2/3} \right) \alpha_D^{z_1,\ldots,z_k}.
\end{equation}

    \item The field $\mathcal E$ formally defined by
\begin{equation}
{\mathcal E}(z) := \frac{\hat{c}}{\sqrt\lambda} \lim_{\varepsilon \to 0} \vartheta_{\varepsilon}^{-1} \big(N_{\varepsilon}(z)-\Braket{N_{\varepsilon}(z)}\big),
\end{equation}
where $N_{\varepsilon}(z)$ counts the number of loops $\ell$ that come to distance $\varepsilon$ of $z$,\footnote{We note that $N_{\varepsilon}(z)$ is infinite with probability one because of the scale invariance of the BLS, but its centered version $E_{\varepsilon}(z) := N_{\varepsilon}(z)-\Braket{N_{\varepsilon}(z)}$ has well defined $n$-point functions---see Lemma \ref{lemma:n-point-function}.} behaves like a conformal primary field with scaling dimension $2/3$. The constant $\hat{c}$ is chosen so that ${\mathcal E}$ is canonically normalized, i.e.
\begin{equation}
\Braket{{\mathcal E}(z_1){\mathcal E}(z_2)}_{\mathbb C} = |z_1-z_2|^{-4/3}.
\end{equation}

    \item More precisely, if $D'$ is a domain conformally equivalent to $D$ and $f:D \to D'$ is a conformal map, then
    \begin{equation}
    \Braket{{\mathcal E}(f(z_1)) \ldots {\mathcal E}(f(z_n))}_{D'} = \left( \prod_{j=1}^n |f'(z_j)|^{-2/3} \right) \Braket{{\mathcal E}(z_1) \ldots {\mathcal E}(z_n)}_{D}.
\end{equation}

    \item Letting $z_{jk} := z_j-z_k$, we have
\begin{align}
    \Braket{\mathcal{O}_\beta(z_1) \mathcal{O}_{-\beta}(z_2) \mathcal{E}(z_3)}_{\mathbb C} = C^{\mathcal{E}}_{\mathcal{O}_\beta \mathcal{O}_{-\beta}} \frac{1}{|z_{12}|^{4\Delta(\beta)}} \left| \frac{z_{12}}{z_{13}z_{23}} \right|^{2/3},
\end{align}
    with \emph{three-point structure constant}
\begin{align}
    C^{\mathcal{E}}_{\mathcal{O}_\beta\mathcal{O}_{-\beta}} = -\sqrt{\lambda}(1-\cos\beta)\frac{2^{7/6} \pi}{3^{1/4} \sqrt{5} \Gamma(1/6)\Gamma(4/3)}.
\end{align}

    \item The OPE of $\mathcal{O}_{\beta} \times \mathcal{O}_{-\beta}$ takes the form
    \begin{align}
    \begin{split}
    & \mathcal{O}_{\beta}(z) \times \mathcal{O}_{-\beta}(z') \\
    & \quad = |z-z'|^{-4\Delta(\beta)} \Big(\mathbb{1} + C^{\mathcal{E}}_{\mathcal{O}_\beta\mathcal{O}_{-\beta}} |z-z'|^{2/3} {\mathcal E}(z)+
    C^{\mathcal{E}^{(2)}}_{\mathcal{O}_\beta\mathcal{O}_{-\beta}} |z-z'|^{4/3} {\mathcal E}^{(2)}(z) \\
    & \qquad + o\big(|z-z'|^{4/3}\big)\Big),
    \end{split}
    \end{align}
    where $\mathbb{1}$ is the identity operator and
    \begin{align}
        \left( C^{\mathcal{E}^{(2)}}_{\mathcal{O}_\beta\mathcal{O}_{-\beta}} \right)^2 =
        \frac{1}{2} \left( C^{\mathcal{E}}_{\mathcal{O}_\beta\mathcal{O}_{-\beta}}\right)^4.
    \end{align}

    \item The mixed full-plane four-point function $\Braket{\mathcal{O}_{\beta}(z_1) \mathcal{O}_{-\beta}(z_2) {\mathcal E}(z_3) {\mathcal E}(z_4)}_{\mathbb C}$ has the following explicit expression:
    \begin{align}
    \begin{split} \label{VVEE_summary}
    & \Braket{\mathcal{O}_{\beta}(z_1) \mathcal{O}_{-\beta}(z_2) {\mathcal E}(z_3) {\mathcal E}(z_4)}_{\mathbb C} \\
    & = |z_{12}|^{-4\Delta(\beta)} \left[ \frac{1+\cos\beta}{2} |z_{34}|^{-4/3} + \frac{1-\cos\beta}{2} Z_{\text{twist}} + \lambda (1-\cos\beta)^2 \hat{\alpha}^{z_3}_{z_1|z_2} \hat{\alpha}^{z_4}_{z_1|z_2} \right],
\end{split}
\end{align}
where
\begin{align}
    \hat{\alpha}^{z_l}_{z_j|z_k;{\mathbb C}} = \frac{2^{7/6} \pi}{3^{1/4} \sqrt{5} \, \Gamma(1/6)\Gamma(4/3)} \left| \frac{z_{jk}}{z_{jl}z_{kl}} \right|^{2/3}
\end{align}
and
\begin{align}
\begin{split}
    & Z_{\text{twist}} = \left| \frac{z_{13} z_{24} }{ z_{34}^2 z_{23} z_{14} } \right|^{2/3} \Bigg[
    \left| {}_2 F_1\left( -\frac{2}{3}, \frac{1}{3}; \frac{2}{3}, \frac{z_{12}z_{34}}{z_{13}z_{24}} \right)  \right|^2 \\
    & \qquad - \frac{4\Gamma\left( \frac{2}{3} \right)^6}{\Gamma\left( \frac{4}{3} \right)^2 \Gamma\left( \frac{1}{3} \right)^4} \left| \frac{z_{12}z_{34}}{z_{13}z_{24}} \right|^{2/3} \left| {}_2 F_1\left( -\frac{1}{3}, \frac{2}{3}; \frac{4}{3}, \frac{z_{12}z_{34}}{z_{13}z_{24}} \right) \right|^2
    \Bigg].
\end{split}
\end{align}

    \item The OPE of $\mathcal{E} \times \mathcal{E}$ contains the terms
\begin{align}
\begin{split}
    & {\mathcal E}(z) \times {\mathcal E}(z') \\
    &= |z-z'|^{-4/3} \left( \mathbb{1} + C^{\mathcal E}_{{\mathcal E}{\mathcal E}} |z-z'|^{2/3} {\mathcal E}(z) +
    C^{\mathcal{E}^{(2)}}_{{\mathcal E}{\mathcal E}} |z-z'|^{4/3} \mathcal{E}^{(2)}(z)
    + \ldots \right),
\end{split}
\end{align}
where the \emph{three-point structure constants} are
\begin{align}
    C_{\mathcal{E}\mathcal{E}}^{\mathcal{E}} &= \frac{1}{\sqrt{\lambda}} \frac{2^{13/6} \, 3^{1/4} \, \sqrt{5} \, \pi^{3/2} \, \Gamma\left(\frac{2}{3}\right)}{\Gamma\left(\frac{1}{6}\right)^3 \Gamma\left(\frac{7}{6}\right)} \\
    C_{\mathcal{E}\mathcal{E}}^{\mathcal{E}^{(2)}} &= \sqrt{2}.
\end{align}

    \item The OPE of $\mathcal{O}_\beta \times \mathcal{E}$ takes the form
\begin{align}
    \mathcal{O}_\beta(z) \times \mathcal{E}(z') = 
    C_{\mathcal{O}_\beta \mathcal{E}}^{\mathcal{O}_\beta} |z-z'|^{-2/3} \mathcal{O}_\beta(z)
    + C_{\mathcal{O}_\beta \mathcal{E}}^{\mathcal{E}_\beta} \mathcal{E}_\beta(z) +  \ldots 
\end{align}
where $C_{\mathcal{O}_\beta \mathcal{E}}^{\mathcal{O}_\beta} = C_{\mathcal{O}_\beta \mathcal{O}_{-\beta}}^{\mathcal{E}}$ and
\begin{align}
    \left( C_{\mathcal{O}_\beta \mathcal{E}}^{\mathcal{E}_\beta}\right)^2 = \frac{1 + \cos\beta}{2}.
\end{align}

    \item The higher-order edge operators $\mathcal{E}^{(k)}$ behave like canonically normalized primary fields. More precisely, for each $k \in \mathbb{N}$,
    \begin{equation}
    \Braket{{\mathcal E}^{(k)}(z_1){\mathcal E}^{(k)}(z_2)}_{\mathbb C} = |z_1-z_2|^{-4k/3}.
    \end{equation}
    Moreover, if $D'$ is a domain conformally equivalent to $D$ and $f:D \to D'$ is a conformal map, then
    \begin{align}
    \begin{split}
    & \Braket{ \mathcal{E}^{(k_1)}(f(z_1)) \ldots \mathcal{E}^{(k_n)}(f(z_n)) }_{D'} \\
    & \qquad = \left( \prod_{j=1}^n |f'(z_j)|^{-2k_j/3} \right) \Braket{ \mathcal{E}^{(k_1)}(z_1) \ldots \mathcal{E}^{(k_n)}(z_n) }_{D}.
    \end{split}
\end{align}

    \item The central charge of the BLS can be independently re-derived to be $c=2\lambda$
    by computing the two-point function of the stress-tensor
    \begin{align}
        \Braket{T(z_1) T(z_2)}_\mathbb{C} = \frac{c/2}{z_{12}^4}
    \end{align}
    from \eqref{VVEE_summary}
    by applying the OPEs of $\mathcal{E}\times\mathcal{E}$ and $\mathcal{O}_{\beta} \times \mathcal{O}_{-\beta}$.
\end{itemize}

\subsection{Structure of the paper}

This paper contains both rigorous results and ``physics-style'' arguments and is written with a mixed audience of mathematicians and physicists in mind. The rigorous results are generally presented as lemmas or theorems in the text; they include explicit expressions for certain correlation functions and the proof that the $n$-point correlation functions of the edge operator $\mathcal E$ and of the higher-order edge operators $\mathcal{E}^{(k)}$ are conformally covariant. The proofs of most rigorous results are collected in the appendix to avoid breaking the flow of the paper. The results in Sections \ref{sec:edge}-\ref{sec:mixed} and \ref{sec:higher_order} are rigorous except for the use of Eq.\ (6.19) of \cite{Camia_2020} in Section \ref{sec:starmeasure}, the existence of the limit in \eqref{def:tildec} in Section \ref{sec:OPE}, the use of Eq.\ (52) of \cite{Simmons_2009} and the identification in \eqref{eq:Ztwist-explained} in Section \ref{sec:mixed}.

The edge operator $\mathcal{E}$ is introduced in Section \ref{sec:edge}, where its correlation functions are discussed. Section \ref{sec:starmeasure} contains the computation of $\Braket{\mathcal{O}_\beta(z_1) \mathcal{O}_{-\beta}(z_2) \mathcal{E}(z_3)}_{\mathbb C}$, including the structure constant $C^{\mathcal{E}}_{\mathcal{O}_\beta \mathcal{O}_{-\beta}}$. Section \ref{sec:OPE} contains a derivation of the OPE of $\mathcal{O}_{\beta} \times \mathcal{O}_{-\beta}$ and the identification of the edge operator $\mathcal{E}$ with the primary operator of dimension $(1/3,1/3)$ discovered in \cite{Camia_2020}. Section \ref{sec:mixed} contains the calculation of the full-plane four-point function $\Braket{\mathcal{O}_\beta(z_1) \mathcal{O}_{-\beta}(z_2) \mathcal{E}(z_3) \mathcal{E}(z_4)}_{\mathbb C}$. Higher-order and charged edge operators are introduced in Sections \ref{sec:higher_order} and \ref{sec:charged}, respectively, where their correlation functions are discussed.
The Virasaoro conformal block expansion resulting from the four-point function $\Braket{\mathcal{O}_\beta(z_1) \mathcal{O}_{-\beta}(z_2) \mathcal{E}(z_3) \mathcal{E}(z_4)}_{\mathbb C}$ is developed in Section \ref{sec:conf-blocks}, while Section \ref{sec:3edges} contains a direct derivation of the full-plane three-point function $\Braket{\mathcal{E}(z_1) \mathcal{E}(z_2) \mathcal{E}(z_3)}_{\mathbb C}$, including the structure constant $C^{\mathcal{E}}_{\mathcal{E} \mathcal{E}}$.
Section \ref{sec:c} contains a new derivation of the fact that the central charge of the BLS with intensity $\lambda$ is $c=2\lambda$.

\section{The edge counting operator} \label{sec:edge}

For a domain $D \subseteq {\mathbb C}$, a point $z \in {\mathbb C}$, a real number $\varepsilon>0$, and a collection $\mathcal L$ of simple loops in $D$, let $n^{\varepsilon}_z(\mathcal L)$ denote the number of loops $\ell \in {\mathcal L}$ such that $\ell \cap B_{\varepsilon}(z) \neq \emptyset$, where $B_{\varepsilon}(z)$ denotes the disk of radius $\varepsilon$ centered at $z$. We define \emph{formally} the ``random variable'' $N_{\varepsilon}(z) = n^{\varepsilon}_z(\mathcal L)$ where $\mathcal L$ is distributed like the collection of outer boundaries $\ell=\ell(\gamma)$ of the loops $\gamma$ of a Brownian loop soup in $D$ with intensity $\lambda$ (see Section \ref{sec:definitions}).

$N_{\varepsilon}(z)$ counts the number of loops $\gamma$ of a Brownian loop soup whose ``edge'' $\ell$ (the outer boundary) comes $\varepsilon-$close to $z$; it is only formally defined because it is infinite with probability one. Nevertheless, we will be interested in the fluctuations of 
$N_{\varepsilon}(z)$ around its infinite mean, which can be formally written as
\begin{align}
\begin{split} \label{Edef}
    E_{\varepsilon}(z) := & N_{\varepsilon}(z) - \langle N_{\varepsilon}(z) \rangle_D \\
    = & N_{\varepsilon}(z) - \lambda\mu^{\text{loop}}_D(\ell \cap B_{\varepsilon}(z) \neq \emptyset),
\end{split}
\end{align}
where $\langle \cdot \rangle_D$ denotes expectation with respect to the Brownian loop soup in $D$ (of fixed intensity $\lambda$) and $\mu^{\text{loop}}_D$ is the Brownian loop measure restricted to $D$, i.e.\ the unique (up to a multiplicative constant) conformally invariant measure on simple planar loops \cite{2005math.....11605W}.

In Lemma \ref{lemma:cutoff-n-point-functions} of the appendix we show that, while $E_{\varepsilon}(z)$ is only formally defined, its correlation functions $\langle E_{\varepsilon}(z_1) \ldots E_{\varepsilon}(z_n) \rangle_D$ are well defined for any collection of points $z_1, \ldots, z_n$ at distance greater than $2\varepsilon$ from each other, with $n \geq 2$. There is a closed-form expression for such correlations in terms of the Brownian loop measure $\mu_D^{\text{loop}}$, as stated in the following lemma, whose proof is presented to the appendix.

\begin{lemma} \label{lemma:n-point-function}
For any $\varepsilon>0$ and any collection of distinct points $z_1,\ldots,z_n \in D$ at distance greater than $2\varepsilon$ from each other, with $n \geq 2$, let $\Pi$ denote the set of all partitions of $\{1,\ldots,n\}$ such that each element $I_l$ of $\{I_1,\ldots,I_r\} \in \Pi$ has cardinality $|I_l| \geq 2$; then
\begin{equation} \label{eq:n-point-function}
    \Braket{E_{\varepsilon}(z_1) \ldots E_{\varepsilon}(z_n)}_D = \sum_{\{I_1,\ldots,I_r\} \in \Pi} \lambda^r \prod_{l=1}^r \mu^{\text{loop}}_D(\ell \cap B_{\varepsilon}(z_j) \neq \emptyset \; \forall j \in I_l).
\end{equation}
\end{lemma}

We remind the reader that $\vartheta_{\varepsilon}$ denotes the scaling limit of the probability that, in critical site percolation on the triangular lattice, there are one open and two closed paths crossing the annulus with inner radius $\varepsilon$ and outer radius $1$, known as a three-arm event, and that $\vartheta_{\varepsilon}\sim\varepsilon^{2/3}$.
A central result of this paper is the fact that the field formally defined by
\begin{equation} \label{def:edge}
{\mathcal E}(z) := \frac{\hat{c}}{\sqrt\lambda} \lim_{\varepsilon \to 0} \vartheta_{\varepsilon}^{-1} E_{\varepsilon}(z)
\end{equation}
behaves like a conformal primary field, where the constant $\hat{c}$ is chosen to ensure that ${\mathcal E}$ is canonically normalized, i.e.,
\begin{equation} \label{eq:canonical_edge}
\Braket{{\mathcal E}(z_1){\mathcal E}(z_2)}_{\mathbb C} = |z_1-z_2|^{-4/3}.
\end{equation}
This result relies crucially on the following lemma, which is interesting in its own right.

\begin{lemma} \label{lemma:conf-cov-mu}
Let $D \subseteq{\mathbb C}$ be either the complex plane $\mathbb C$ or the upper-half plane $\mathbb H$ or any domain conformally equivalent to $\mathbb H$.
For any collection of distinct points $z_1,\ldots,z_k \in D$ with $k \geq 2$, the following limit exists:
\begin{equation} \label{eq:epsilon-limit}
    \alpha_D^{z_1,\ldots,z_k} := \lim_{\varepsilon \to 0} \vartheta_{\varepsilon}^{-k} \mu^{\text{loop}}_D(\ell \cap B_{\varepsilon}(z_j) \neq \emptyset \; \forall j=1,\ldots,k).
\end{equation}
Moreover, $\alpha_D^{z_1,\ldots,z_k}$ is conformally covariant in the sense that, if $D'$ is a domain conformally equivalent to $D$ and $f:D \to D'$ is a conformal map, then
\begin{equation} \label{eq:conf-cov-mu}
    \alpha_{D'}^{f(z_1),\ldots,f(z_k)} = \left( \prod_{j=1}^k |f'(z_j)|^{-2/3} \right) \alpha_D^{z_1,\ldots,z_k}.
\end{equation}
\end{lemma}

For any collection of points $z_1,\ldots,z_n \in D$ and any subset $S=\{z_{j_1},\ldots,z_{j_k}\}$ of $\{z_1,\ldots,z_n\}$, let $\alpha^S_D := \alpha_D^{z_{j_1},\ldots,z_{j_k}}$. The statement about the operator $\mathcal E$ defined formally in \eqref{def:edge} is made precise by the following theorem.

\begin{theorem} \label{thm:edge}
Let $D \subseteq{\mathbb C}$ be either the complex plane $\mathbb C$ or the upper-half plane $\mathbb H$ or any domain conformally equivalent to $\mathbb H$. For any collection of distinct points $z_1, \ldots, z_n \in D$ with $n \geq 2$, the following limit exists:
\begin{equation} \label{eq:existence}
g_D(z_1, \ldots, z_n) := \lim_{\varepsilon \to 0} \vartheta_{\varepsilon}^{-n} \Braket{ E_{\varepsilon}(z_1) \ldots E_{\varepsilon}(z_n) }_D.
\end{equation}
Moreover, if $\mathcal S = \mathcal{S}(z_1,\ldots,z_n)$ denotes the set of all partitions of $\{z_1,\ldots,z_n\}$ such that each element $S_l$ of $(S_1,\ldots,S_r) \in \mathcal{S}$ has cardinality $|S_l| \geq 2$, then
\begin{equation} \label{eq:g}
g_D(z_1, \ldots, z_n) = \sum_{(S_1,\ldots,S_r) \in \mathcal{S}} \lambda^r \alpha^{S_1}_D\ldots\alpha^{S_r}_D.
\end{equation}
Furthermore, $g_D(z_1, \ldots, z_n)$ is conformally covariant in the sense that, if $D'$ is a domain conformally equivalent to $D$ and $f:D \to D'$ is a conformal map, then
\begin{equation} \label{eq:conf-cov}
    g_{D'}(f(z_1), \ldots, f(z_n)) = \left( \prod_{k=1}^n |f'(z_k)|^{-2/3} \right) g_D(z_1, \ldots, z_n).
\end{equation}
\end{theorem}

\noindent{\bf Proof.} The existence of the limit in \eqref{eq:existence} follows from \eqref{eq:n-point-function} combined with the existence of the limit in \eqref{eq:epsilon-limit}. The expression in \eqref{eq:g} follows directly from \eqref{eq:n-point-function} and the definition of $\alpha^{z_1,\ldots,z_k}(D)$ in \eqref{eq:epsilon-limit}. The conformal covariance expressed in \eqref{eq:conf-cov} is an immediate consequence of \eqref{eq:g} and \eqref{eq:conf-cov-mu}. \qed

\medskip

Using the notation introduced in \eqref{def:edge}, we will write
\begin{equation} \label{def:n-point-function}
    \langle {\mathcal E}(z_1) \ldots {\mathcal E}(z_n) \rangle_D := \frac{\hat{c}^n}{\lambda^{n/2}} g_D(z_1, \ldots, z_n),
\end{equation}
despite the fact that ${\mathcal E}$ is only formally defined.
To simplify the notation, we define
\begin{equation} \label{def:alphahat}
    \hat{\alpha}^{z_1,\ldots,z_k}_{D} := \hat{c}^k \, {\alpha}^{z_1,\ldots,z_k}_{D}.
\end{equation}
In particular, using this notation, the two-, three- and four-point functions are
\begin{subequations}
\begin{align}
\Braket{{\mathcal E}(z_1) {\mathcal E}(z_2)}_D & = \hat{\alpha}_D^{z_1,z_2} \label{eq:EE-alpha} \\
\Braket{{\mathcal E}(z_1) {\mathcal E}(z_2) {\mathcal E}(z_3)}_D & = \frac{1}{\sqrt\lambda}\hat{\alpha}_D^{z_1,z_2,z_3} \label{eq:EEE-alpha} \\
\Braket{{\mathcal E}(z_1) {\mathcal E}(z_2) {\mathcal E}(z_3) {\mathcal E}(z_4)}_D & = \frac{1}{\lambda} \hat{\alpha}_D^{z_1,z_2,z_3,z_4} \nonumber \\
& \quad + \hat{\alpha}_D^{z_1,z_2}\hat{\alpha}_D^{z_3,z_4} + \hat{\alpha}_D^{z_1,z_3}\hat{\alpha}_D^{z_2,z_4} + \hat{\alpha}_D^{z_1,z_4}\hat{\alpha}_D^{z_2,z_3}. \label{eq:EEEE-alpha}
\end{align}
\end{subequations}

\section{Correlation functions with a ``twist''} \label{sec:starmeasure}

In this section we present a simple method to compute certain types of correlation functions involving two vertex layering operators. Later, as an application, we will use this method to show how the edge operator $\mathcal E$ emerges from the OPE of $\mathcal{O}_{\beta} \times \mathcal{O}_{-\beta}$. From now on, we will drop the subscript $D$ from $\Braket{\cdot}_D$, $\mu^{\text{loop}}_D$, $\alpha_D^{z_1,\ldots,z_k}$ and similar expressions when $D$ can be any domain. 

To explain the method mentioned above, in the next paragraph we use $\{ \cdot \}$ to denote an unnormilazed sum, that is
\begin{equation}
\Braket{ \cdot } := \frac{1}{Z} \{ \cdot \},
\end{equation}
where $Z := \{ 1 \}$ denotes the partition function. If we define
\begin{align}
    \{ \cdot \}^*_{z_1,z_2} \equiv \{ \cdot \}^*_{z_1,z_2;\beta} := \{ \cdot \; \mathcal{O}_{\beta}(z_1) \mathcal{O}_{-\beta}(z_2) \}
\end{align}
and
\begin{align}
    \Braket{\cdot}^*_{z_1,z_2} \equiv \Braket{\cdot}^*_{z_1,z_2;\beta} := \frac{\{ \cdot \}^*_{z_1,z_2}}{\{ 1 \}^*_{z_1,z_2} },
\end{align}
then we can write
\begin{align}
\begin{split}
    \Braket{\cdot \; \mathcal{O}_{\beta}(z_1) \mathcal{O}_{-\beta}(z_2)} & = \frac{\{ \cdot \; \mathcal{O}_{\beta}(z_1) \mathcal{O}_{-\beta}(z_2) \}}{\{ 1 \} } \\
    & = \frac{\{ 1 \}^*_{z_1,z_2}}{\{ 1 \} } \frac{\{ \cdot \}^*_{z_1,z_2}}{\{ 1 \}^*_{z_1,z_2}} \\
    & = \Braket{\mathcal{O}_{\beta}(z_1) \mathcal{O}_{-\beta}(z_2)} \Braket{\cdot}^*_{z_1,z_2}.
\end{split}
\end{align}
This simple formula will be very useful in the rest of the paper thanks to the observation that $\Braket{\cdot}^*_{z_1,z_2}$ is the expectation with respect to the measure $\mu^*_{z_1,z_2;\beta} \equiv \mu^*_{z_1,z_2}$ defined by
\begin{equation} \label{mu-star}
\mu^*_{z_1,z_2}(\ell) := \left\{
\begin{array}{lll}
    \mu^{\text{loop}}(\ell) & \mbox{if $\ell$ does not separate $z_1,z_2$} \\
    e^{i\beta\sigma_{\ell}} \mu^{\text{loop}}(\ell) & \mbox{if $z_1 \in \bar{\ell}, z_2 \notin \bar{\ell}$} \\
    e^{-i\beta\sigma_{\ell}} \mu^{\text{loop}}(\ell) & \mbox{if $z_1 \notin \bar{\ell}, z_2 \in \bar{\ell}$}
\end{array} \right.
\end{equation}
where $\sigma_{\ell} = \pm 1$ is a symmetric Boolean variable assigned to $\ell$.

As a first example, to illustrate the use of the method, we calculate
\begin{align}
\begin{split}
    \Braket{\mathcal{O}_{\beta}(z_1) \mathcal{O}_{-\beta}(z_2) {\mathcal E}(z_3)} & = \frac{\hat{c}}{\sqrt\lambda} \lim_{\varepsilon \to 0} \vartheta_{\varepsilon}^{-1} \Braket{\mathcal{O}_{\beta}(z_1) \mathcal{O}_{-\beta}(z_2) E_{\varepsilon}(z_3)} \\
    & = \frac{\hat{c}}{\sqrt\lambda} \Braket{\mathcal{O}_{\beta}(z_1) \mathcal{O}_{-\beta}(z_2)} \lim_{\varepsilon \to 0} \vartheta_{\varepsilon}^{-1} \Braket{E_{\varepsilon}(z_3)}^*_{z_1,z_2}.
\end{split}
\end{align}
To perform this calculation, we define $N^{\delta}_{\varepsilon}(z) := n^{\varepsilon}_z({\mathcal L}^{\delta})$ and $E^{\delta}_{\varepsilon}(z) := N^{\delta}_{\varepsilon}(z) - \Braket{N^{\delta}_{\varepsilon}(z)}$, where ${\mathcal L}^{\delta}$ is a Brownian loop soup with cutoff $\delta>0$, obtained by taking the usual Brownian loop soup and removing all loops with diameter (defined to be the largest distance between any two points on the loop) smaller than $\delta$. The random variables $N^{\delta}_{\varepsilon}(z)$ and $E^{\delta}_{\varepsilon}(z)$ are well defined because of the cutoffs $\varepsilon$ and $\delta$. With these definitions, we have
\begin{align}
\begin{split} \label{eq:E^*}
    \Braket{E_{\varepsilon}(z_3)}^*_{z_1,z_2} & := \lim_{\delta \to 0}\Braket{E^{\delta}_{\varepsilon}(z_3)}^*_{z_1,z_2} = \lim_{\delta \to 0} \Big[\Braket{N^{\delta}_{\varepsilon}(z_3)}^*_{z_1,z_2} - \Braket{N^{\delta}_{\varepsilon}(z_3)}\Big] \\
    & = \lim_{\delta \to 0} \left[ (\cos\beta - 1) \lambda \mu^{\text{loop}}(\operatorname{diam}(\ell)>\delta, \ell \cap B_{\varepsilon}(z_3) \neq \emptyset, \ell \text{ separates } z_1, z_2) \right] \\
    & = -\lambda(1-\cos\beta) \mu^{\text{loop}}(\ell \cap B_{\varepsilon}(z_3) \neq \emptyset, \ell \text{ separates } z_1,z_2).
\end{split}
\end{align}
The expression above for $\Braket{E_{\varepsilon}(z_3)}^*_{z_1,z_2}$ follows from the observation that the contributions to $\Braket{N^{\delta}_{\varepsilon}(z_3)}^*_{z_1,z_2}$ and $\Braket{N^{\delta}_{\varepsilon}(z_3)}$ from loops that do not separate $z_1$ and $z_2$ cancel out, while the contribution to $\Braket{N^{\delta}_{\varepsilon}(z_3)}^*_{z_1,z_2}$ from loops that do separate $z_1$ and $z_2$ comes with a factor $\cos\beta$ because of the definition of $\mu^*_{z_1,z_2}$ and the averaging over $\sigma_{\ell}=\pm 1$. (Note that $\{ \sigma_{\ell} \}_{\ell \in {\mathcal L}}$ is distributed like a collection of independent, $(\pm 1)-$valued, symmetric random variables).

We conclude that
\begin{equation} \label{eq:OOE1}
    \Braket{\mathcal{O}_{\beta}(z_1) \mathcal{O}_{-\beta}(z_2) {\mathcal E}(z_3)} = -\sqrt{\lambda} (1-\cos\beta) \, \hat{\alpha}^{z_3}_{z_1|z_2} \Braket{\mathcal{O}_{\beta}(z_1) \mathcal{O}_{-\beta}(z_2)},
\end{equation}
where
\begin{equation} \label{def:alphahat312}
    \hat{\alpha}^{z_3}_{z_1|z_2} := \hat{c} \, {\alpha}^{z_3}_{z_1|z_2}
\end{equation}
with
\begin{equation} \label{def:alpha312}
    \alpha^{z_3}_{z_1|z_2} \equiv \alpha^{z_3}_{z_2|z_1} := \lim_{\varepsilon \to 0} \vartheta_{\varepsilon}^{-1} \mu^{\text{loop}}(\ell \cap B_{\varepsilon}(z_3) \neq \emptyset, \ell \text{ separates } z_1,z_2).
\end{equation}
The existence of the limit in \eqref{def:alpha312} follows from the proof of Lemma \ref{lemma:conf-cov-mu}.

So far our discussion has been completely general and independent of the domain $D$. If we now specify that $D={\mathbb C}$ and note that the operators $\mathcal{O}_{\beta}, \mathcal{O}_{-\beta}$ are canonically normalized
\begin{equation} \label{eq:OO}
    \Braket{\mathcal{O}_{\beta}(z_1)\mathcal{O}_{-\beta}(z_2)}_{\mathbb C} = |z_1-z_2|^{-4\Delta(\beta)},
\end{equation}
we get from \eqref{eq:OOE1}
\begin{equation} \label{eq:OOE2}
    \Braket{\mathcal{O}_{\beta}(z_1) \mathcal{O}_{-\beta}(z_2) {\mathcal E}(z_3)}_{\mathbb C} = -\sqrt{\lambda} (1-\cos\beta) \, \hat{\alpha}^{z_3}_{z_1|z_2;{\mathbb C}} |z_1-z_2|^{-4\Delta(\beta)}.
\end{equation}

Since \eqref{eq:OOE2} is a three-point function of primary operators defined on the full plane, its form is fixed by global conformal invariance up to a multiplicative constant (see, for example, the proof of Theorem 4.5 of \cite{Camia_2016}). In this case, letting $z_{jk} := z_j-z_k$, we have
\begin{align} \label{eq:OOE3}
    \Braket{\mathcal{O}_\beta(z_1) \mathcal{O}_{-\beta}(z_2) \mathcal{E}(z_3)}_{\mathbb C} = C^{\mathcal{E}}_{\mathcal{O}_\beta \mathcal{O}_{-\beta}} \frac{1}{|z_{12}|^{4\Delta(\beta)}} \left| \frac{z_{12}}{z_{13}z_{23}} \right|^{2/3}.
\end{align}
The coefficient $C^{\mathcal{E}}_{\mathcal{O}_{\beta} \mathcal{O}_{-\beta}}$, evaluated at $\beta_1=\beta_2=\pi$, was determined in \cite{Camia_2020}, where it was called $C^{(1,1)}$. Comparing \eqref{eq:OOE2} with \eqref{eq:OOE3} and using the expression for $C^{(1,1)}$ from Eq.\ (6.19) of \cite{Camia_2020} shows that
\begin{align} \label{vvee_weights1}
    \hat{\alpha}^{z_3}_{z_1|z_2;{\mathbb C}} = \frac{ 2^{7/6} \pi}{3^{1/4} \sqrt{5} \, \Gamma(1/6)\Gamma(4/3)} \left| \frac{z_{12}}{z_{13}z_{23}} \right|^{2/3}.
\end{align}
Together with \eqref{eq:OOE2}, this implies that, for general values of $\beta$, we have the three-point function coefficient
\begin{align} \label{eq:C-EOO}
    C^{\mathcal{E}}_{\mathcal{O}_\beta\mathcal{O}_{-\beta}} = -\sqrt{\lambda}(1-\cos\beta)\frac{ 2^{7/6} \pi}{3^{1/4} \sqrt{5} \, \Gamma(1/6) \Gamma(4/3)}.
\end{align}

\section{OPE and the edge operator} \label{sec:OPE}

In this section, applying the method presented in the previous section, we show how the edge operator $\mathcal E$ emerges from the Operator Product Expansion (OPE) of $\mathcal{O}_{\beta} \times \mathcal{O}_{-\beta}$. It is shown in \cite{Camia_2020} that the OPE of the product of two vertex operators $\mathcal{O}_{\beta_i} \times \mathcal{O}_{\beta_j}$ contains operators of dimensions $(\Delta_{ij} + \frac{k}{3}, \Delta_{ij} + \frac{k'}{3})$ for non-negative integers $k, k'$, where $\Delta_{ij} = \frac{\lambda}{10}(1-\cos(\beta_i + \beta_j))$. In what follows, we identify the operator of dimensions $(\frac{1}{3},\frac{1}{3})$ with the edge operator $\mathcal{E}$. 

If $N^{\delta}(z)$ denotes the number of loops of diameter larger than $\delta$ that contain $z$ in their interior, it was shown in \cite{Camia_2016} that the two-point function
\begin{align}
\begin{split} \label{eq:O-2-point-function}
    \Braket{\mathcal{O}_{\beta}(z)\mathcal{O}_{-\beta}(z')} &\propto \lim_{\delta \to 0} \delta^{-2\Delta(\beta)} \Braket{e^{i \beta N^{\delta}(z)} e^{-i \beta N^{\delta}(z')}} \\
    &= \lim_{\delta \to 0} \delta^{-2\Delta(\beta)} \exp{\big(-\lambda(1-\cos\beta)\mu^{\text{loop}}\big(\ell \text{ separates } z,z', \text{diam}(\ell)>\delta\big)\big)}
\end{split}
\end{align}
exists.

We are interested in the sub-leading behavior of $\mathcal{O}_{\beta}(z) \times \mathcal{O}_{-\beta}(z')$ when $z' \to z$. The two-point function $\Braket{\mathcal{O}_{\beta}(z)\mathcal{O}_{-\beta}(z')}$ diverges when $z' \to z$ (see \eqref{eq:OO}), so we normalize $\mathcal{O}_{\beta}(z) \mathcal{O}_{-\beta}(z')$ by its expectation. Taking two distinct points $z_1,z_2 \neq z,z'$, we compute the four-point function
\begin{equation} \label{eq:starting-point}
\Braket{\frac{\mathcal{O}_{\beta}(z)\mathcal{O}_{-\beta}(z')}{\Braket{\mathcal{O}_{\beta}(z)\mathcal{O}_{-\beta}(z')}}\mathcal{O}_{\beta'}(z_1)\mathcal{O}_{-\beta'}(z_2)} = \Braket{\mathcal{O}_{\beta'}(z_1)\mathcal{O}_{-\beta'}(z_2)} \frac{\Braket{\mathcal{O}_{\beta}(z)\mathcal{O}_{-\beta}(z')}^*_{z_1,z_2;\beta'}}{\Braket{\mathcal{O}_{\beta}(z)\mathcal{O}_{-\beta}(z')}}.
\end{equation}
The loops that do not separate $z_1$ and $z_2$ contribute equally to $\Braket{\mathcal{O}_{\beta}(z)\mathcal{O}_{-\beta}(z')}^*_{z_1,z_2}$ and $\Braket{\mathcal{O}_{\beta}(z)\mathcal{O}_{-\beta}(z')}$, so their contributions cancel out in the ratio on the right-hand side. The loops that do separate $z_1, z_2$ contribute differently, as we have already seen in the computation leading to \eqref{eq:OOE1}. An analogous computation using \eqref{eq:O-2-point-function} gives
\begin{eqnarray} \label{eq:ratio1}
    \frac{\Braket{\mathcal{O}_{\beta}(z)\mathcal{O}_{-\beta}(z')}^*_{z_1,z_2;\beta'}}{\Braket{\mathcal{O}_{\beta}(z)\mathcal{O}_{-\beta}(z')}} & = & \exp{ \left[ (1-\cos\beta') \lambda(1-\cos\beta) \mu^{\text{loop}}(\ell \text{ separates } z,z' \text{ and } z_1,z_2) \right]} \nonumber \\
    & = & 1 + (1-\cos\beta') \lambda(1-\cos\beta) \mu^{\text{loop}}(\ell \text{ separates } z,z' \text{ and } z_1,z_2) \nonumber \\
    & & + \, O(\mu^{\text{loop}}(\ell \text{ separates } z,z' \text{ and } z_1,z_2)^{2}),
\end{eqnarray}
as $|z-z'| \to 0$.

We now let $\varepsilon=|z-z'|$ and observe that
\begin{align}
\begin{split} \label{eq:tilde-c-epsilon-mu}
    & \mu^{\text{loop}}(\ell \text{ separates } z,z' \text{ and } z_1,z_2) = \mu^{\text{loop}}(\ell \cap B_{\varepsilon}(z) \neq \emptyset \text{ and } \ell \text{ separates } z_1,z_2) \\
    &- \mu^{\text{loop}}(\ell \cap B_{\varepsilon}(z) \neq \emptyset, \ell \text{ does not separate } z,z' \text{ and } \ell \text{ separates } z_1,z_2) \\ & = \mu^{\text{loop}}(\ell \cap B_{\varepsilon}(z) \neq \emptyset \text{ and } \ell \text{ separates } z_1,z_2) \\
    & \quad \left[1 - \frac{\mu^{\text{loop}}(\ell \cap B_{\varepsilon}(z) \neq \emptyset, \ell \text{ does not separate } z,z' \text{ and } \ell \text{ separates } z_1,z_2)}{\mu^{\text{loop}}(\ell \cap B_{\varepsilon}(z) \neq \emptyset \text{ and } \ell \text{ separates } z_1,z_2)} \right],
\end{split}
\end{align}
where
\begin{equation} \label{eq:O1}
    \mu^{\text{loop}}(\ell \cap B_{\varepsilon}(z) \neq \emptyset \text{ and } \ell \text{ separates } z_1,z_2) = O\big(\vartheta_{\varepsilon}\big) \; \text{ as } \varepsilon \to 0,
\end{equation}
which follows from the proof of Lemma \ref{lemma:conf-cov-mu}. Letting
\begin{align} \label{def:tilde-c-epsilon}
    & \tilde{c}_{\varepsilon} \equiv \tilde{c}_{\varepsilon}(z,z';z_1,z_2) \nonumber \\
    & \quad := 1 - \frac{\mu^{\text{loop}}(\ell \cap B_{\varepsilon}(z) \neq \emptyset, \ell \text{ does not separate } z,z' \text{ and } \ell \text{ separates } z_1,z_2)}{\mu^{\text{loop}}(\ell \cap B_{\varepsilon}(z) \neq \emptyset \text{ and } \ell \text{ separates } z_1,z_2)}
\end{align}
and using \eqref{eq:ratio1}-\eqref{def:tilde-c-epsilon}, \eqref{eq:E^*}, and the fact that $\vartheta_{\varepsilon}\sim\varepsilon^{2/3}$, we can write
\begin{equation} \label{eq:ratio3}
\frac{\Braket{\mathcal{O}_{\beta}(z)\mathcal{O}_{-\beta}(z')}^*_{z_1,z_2;\beta'}}{\Braket{\mathcal{O}_{\beta}(z)\mathcal{O}_{-\beta}(z')}} = 1 - (1-\cos\beta) \, \tilde{c}_{\varepsilon} \Braket{E_{\varepsilon}(z)}^*_{z_1,z_2;\beta'} + \, o\big(\varepsilon^{2/3}\big) \; \text{ as } \varepsilon \to 0.
\end{equation}
Combining this with \eqref{eq:starting-point}, we obtain
\begin{align}
\begin{split}
& \Braket{\frac{\mathcal{O}_{\beta}(z)\mathcal{O}_{-\beta}(z')}{\Braket{\mathcal{O}_{\beta}(z_)\mathcal{O}_{-\beta}(z')}}\mathcal{O}_{\beta'}(z_1)\mathcal{O}_{-\beta'}(z_2)} \\
&= \Braket{\mathcal{O}_{\beta'}(z_1)\mathcal{O}_{-\beta'}(z_2)}
- (1-\cos\beta) \, \tilde{c}_{\varepsilon} \, \Braket{\mathcal{O}_{\beta'}(z_1)\mathcal{O}_{-\beta'}(z_2) E_{\varepsilon}(z)} + o\big(\varepsilon^{2/3}\big)
\end{split}
\end{align}
as $\varepsilon \to 0$.

At this point we make the natural assumption that, as long as the points $z, z_1, z_2$ are distinct, the limit
\begin{equation} \label{def:tildec}
    \tilde{c} := \lim_{z' \to z} \tilde{c}_{\varepsilon} \equiv \lim_{z' \to z} \tilde{c}_{\varepsilon}(z,z';z_1,z_2)
\end{equation}
exists and is independent of the domain and of $z,z_1,z_2$. This can be justified using arguments analogous to those in the proof of Lemma \ref{lemma:conf-cov-mu}. The idea is, essentially, the following. One can think in terms of the full scaling limit of critical percolation, as described in the proof of Lemma \ref{lemma:conf-cov-mu}. Then one can split the loops separating $z_1,z_2$ and intersecting $B_{\varepsilon}(z)$ into excursions from $\partial B_{\varepsilon}(z)$ either inside or outside the disk. As explained in the proof of Lemma \ref{lemma:conf-cov-mu}, the excursions inside and outside $B_{\varepsilon}(z)$ are independent of each other, conditioned on the location on $\partial B_{\varepsilon}(z)$ of their starting and ending points. Since the limit in \eqref{def:tildec} is determined only by the behavior of the excursions inside $B_{\varepsilon}(z)$, it should not depend on the domain and on $z_1,z_2$.

Using the assumption expressed by \eqref{def:tildec} and the formal definition \eqref{def:edge} of the edge operator, we can write
\begin{align}
    \frac{\mathcal{O}_{\beta}(z)\mathcal{O}_{-\beta}(z')}{\Braket{\mathcal{O}_{\beta}(z)\mathcal{O}_{-\beta}(z')}} = \mathbb{1} - (1-\cos\beta) \, \frac{\tilde{c}}{\hat{c}} \, \sqrt{\lambda} |z-z'|^{2/3} {\mathcal E}(z) + o\big(|z-z'|^{2/3}\big) \; \text{ as } z' \to z,
\end{align}
where $\mathbb{1}$ denotes the identity operator. For $z$ away from any boundary and in the limit $z' \to z$, using \eqref{eq:OO} this takes the form
\begin{align} \label{eq:OPE}
    \mathcal{O}_{\beta}(z) \times \mathcal{O}_{-\beta}(z') = |z-z'|^{-4\Delta(\beta)} \left(\mathbb{1} - \sqrt{\lambda} (1-\cos\beta) \, \frac{\tilde{c}}{\hat{c}} \, |z-z'|^{2/3} {\mathcal E}(z)+ o\big(|z-z'|^{2/3}\big)\right),
\end{align}
which shows how the edge operator emerges from the OPE of two layering vertex operators.

In order to check for internal consistency, we determine $\tilde{c}/\hat{c}$. To do this we insert the OPE \eqref{eq:OPE} in the three-point function
\begin{align} \label{eq:OOE4}
\begin{split}
    & \Braket{\mathcal{O}_\beta(z_1) \mathcal{O}_{-\beta}(z_2) \mathcal{E}(z_3)}_{\mathbb C} \\
    & \quad = |z_{12}|^{-4\Delta(\beta)} \left( -\sqrt\lambda (1-\cos\beta) \, \frac{\tilde{c}}{\hat{c}} \, \Braket{{\mathcal E}(z_1){\mathcal E}(z_3)}_{\mathbb C} \, |z_{12}|^{2/3} + o\big(|z_{12}|^{2/3}\big) \right).
\end{split}
\end{align}
Comparing this with \eqref{eq:OOE2} and using \eqref{vvee_weights1} and the fact that $\mathcal E$ is assumed to be canonically normalized, so that
\begin{equation} \label{eq:EE}
    \Braket{\mathcal{E}(z_1)\mathcal{E}(z_3)}_{\mathbb C} = |z_{13}|^{-4/3},
\end{equation}
we get
\begin{align}
\begin{split}
  \frac{\tilde{c}}{\hat{c}} \, |z_{13}|^{-4/3} \, |z_{12}|^{2/3} + o\big(|z_{12}|^{2/3}\big) & = \hat{\alpha}^{z_3}_{z_1|z_2;{\mathbb C}} \\
    & = \frac{2^{7/6} \pi}{3^{1/4} \sqrt{5} \, \Gamma(1/6)\Gamma(4/3)} \left| \frac{z_{12}}{z_{13}z_{23}} \right|^{2/3}.
\end{split}
\end{align}
Dividing both sides of the equation above by $|z_{12}|^{2/3}$ and letting $z_2 \to z_1$ gives
\begin{equation} \label{eq:tildec}
   \frac{\tilde{c}}{\hat{c}} = \frac{2^{7/6} \pi}{3^{1/4} \sqrt{5} \, \Gamma(1/6)\Gamma(4/3)}.
\end{equation}

Based on general principles and on the conformal block expansion performed in \cite{Camia_2020}, the OPE of $\mathcal{O}_{\beta} \times \mathcal{O}_{-\beta}$ should read
\begin{align} \label{eq:generalOPE}
    \mathcal{O}_\beta(z) \times \mathcal{O}_{-\beta}(z') = |z-z'|^{-4 \Delta(\beta)} \big(\mathbb{1} + C_{\mathcal{O}_\beta \mathcal{O}_{-\beta}}^{\phi_{1/3,1/3}} |z-z'|^{2/3} \phi_{1/3,1/3}(z) + \ldots\big),
\end{align}
where $\phi_{1/3,1/3}$ is an operator of dimension $(1/3,1/3)$. In order to identify $\phi_{1/3,1/3}$ with the edge operator $\mathcal E$, we need to identify $C_{\mathcal{O}_\beta \mathcal{O}_{-\beta}}^{\phi_{1/3,1/3}}$ with the coefficient $C_{\mathcal{O}_\beta \mathcal{O}_{-\beta}}^{\mathcal E}$ given in \eqref{eq:C-EOO}. Comparing \eqref{eq:generalOPE} with \eqref{eq:OPE}, and using \eqref{eq:tildec}, this gives
\begin{equation}
    C_{\mathcal{O}_\beta \mathcal{O}_{-\beta}}^{\phi_{1/3,1/3}} = -\sqrt{\lambda} (1-\cos\beta) \frac{2^{7/6} \pi}{3^{1/4} \sqrt{5} \, \Gamma(1/6)\Gamma(4/3)},
\end{equation}
which indeed coincides with \eqref{eq:C-EOO}.

\section{A mixed four-point function} \label{sec:mixed}

The method introduced in Section \ref{sec:starmeasure} can be used to calculate the mixed four-point function
\begin{align}
\begin{split} \label{eq:OOEE-start}
    \Braket{\mathcal{O}_{\beta}(z_1) \mathcal{O}_{-\beta}(z_2) {\mathcal E}(z_3) {\mathcal E}(z_4)} & = \Braket{\mathcal{O}_{\beta}(z_1) \mathcal{O}_{-\beta}(z_2)} \Braket{{\mathcal E}(z_3) {\mathcal E}(z_4)}^*_{z_1,z_2} \\
    &= \lambda^{-1} \hat{c}^2 \, \Braket{\mathcal{O}_{\beta}(z_1) \mathcal{O}_{-\beta}(z_2)} \lim_{\varepsilon \to 0} \vartheta_{\varepsilon}^{-2} \Braket{E_{\varepsilon}(z_3) E_{\varepsilon}(z_4)}^*_{z_1,z_2}.
\end{split}
\end{align}
Using the random variables defined just above \eqref{eq:E^*}, a bit of algebra shows that
\begin{align}
\begin{split}
    & \Braket{E_{\varepsilon}(z_3) E_{\varepsilon}(z_4)}^*_{z_1,z_2} = \lim_{\delta \to 0} \Braket{E^{\delta}_{\varepsilon}(z_3) E^{\delta}_{\varepsilon}(z_4)}^*_{z_1,z_2} \\
    &= \lim_{\delta \to 0} \Braket{\Big[N^{\delta}_{\varepsilon}(z_3) - \Braket{N^{\delta}_{\varepsilon}(z_3)}\Big] \Big[N^{\delta}_{\varepsilon}(z_4) - \Braket{N^{\delta}_{\varepsilon}(z_4)}\Big]}^*_{z_1,z_2} \\
    &= \lim_{\delta \to 0} \Braket{\Big[N^{\delta}_{\varepsilon}(z_3) - \Braket{N^{\delta}_{\varepsilon}(z_3)}^*_{z_1,z_2}\Big] \Big[N^{\delta}_{\varepsilon}(z_4) - \Braket{N^{\delta}_{\varepsilon}(z_4)}^*_{z_1,z_2}\Big]}^*_{z_1,z_2} \\
    & \quad + \Braket{E_{\varepsilon}(z_3)}^*_{z_1,z_2} \Braket{E_{\varepsilon}(z_4)}^*_{z_1,z_2}.
\end{split}
\end{align}
Now note that
\begin{align}
    \lim_{\delta \to 0} \Braket{\Big[N^{\delta}_{\varepsilon}(z_3) - \Braket{N^{\delta}_{\varepsilon}(z_3)}^*_{z_1,z_2}\Big] \Big[N^{\delta}_{\varepsilon}(z_4) - \Braket{N^{\delta}_{\varepsilon}(z_4)}^*_{z_1,z_2}\Big]}^*_{z_1,z_2}
\end{align}
is exactly analogous to $\Braket{E_{\varepsilon}(z_3) E_{\varepsilon}(z_3)}$, with the measure $\mu^{\text{loop}}$ replaced by $\mu^*_{z_1,z_2}$. Therefore, combining Lemma \ref{lemma:n-point-function} with \eqref{mu-star}, we have that
\begin{align}
\begin{split}
    & \lim_{\delta \to 0} \Braket{\Big[N^{\delta}_{\varepsilon}(z_3) - \Braket{N^{\delta}_{\varepsilon}(z_3)}^*_{z_1,z_2}\Big] \Big[N^{\delta}_{\varepsilon}(z_4) - \Braket{N^{\delta}_{\varepsilon}(z_4)}^*_{z_1,z_2}\Big]}^*_{z_1,z_2} \\
    & = \lambda \mu^*_{z_1,z_2}(\ell \cap B_{\varepsilon}(z_j) \neq \emptyset \; \text{ for } j=3,4) \\
    & = \lambda \mu^{\text{loop}}(\ell \cap B_{\varepsilon}(z_j) \neq \emptyset \; \text{ for } j=3,4; \ell \text{ does not separate } z_1,z_2) \\
    & \quad + \lambda \cos\beta \mu^{\text{loop}}(\ell \cap B_{\varepsilon}(z_j) \neq \emptyset \; \text{ for } j=3,4; \ell \text{ separates } z_1,z_2) \\
    & = \lambda \mu^{\text{loop}}(\ell \cap B_{\varepsilon}(z_j) \neq \emptyset \; \text{ for } j=3,4) \\
    & \quad - \lambda (1 - \cos\beta) \mu^{\text{loop}}(\ell \cap B_{\varepsilon}(z_j) \neq \emptyset \; \text{ for } j=3,4; \ell \text{ separates } z_1,z_2).
\end{split}
\end{align}
Using this and \eqref{eq:E^*}, we obtain
\begin{align} \label{eq:EE^*}
\begin{split}
    &\Braket{E_{\varepsilon}(z_3) E_{\varepsilon}(z_4)}^*_{z_1,z_2}
    = \lambda \mu^{\text{loop}}(\ell \cap B_{\varepsilon}(z_j) \neq \emptyset \text{ for } j=3,4) \\
    & \quad - \lambda (1-\cos\beta) \mu^{\text{loop}}(\ell \cap B_{\varepsilon}(z_j) \neq \emptyset \text{ for } j=3,4; \ell \text{ separates } z_1,z_2) \\
    & \quad + \lambda^2 (1-\cos\beta)^2 \mu^{\text{loop}}(\ell \cap B_{\varepsilon}(z_3) \neq \emptyset, \ell \text{ separates } z_1,z_2) \\
    & \qquad \cdot  \mu^{\text{loop}}(\ell \cap B_{\varepsilon}(z_4) \neq \emptyset, \ell \text{ separates } z_1,z_2).
\end{split}
\end{align}
Inserting this expression in \eqref{eq:OOEE-start} gives
\begin{align} \label{eq:OOEE}
    & \Braket{\mathcal{O}_{\beta}(z_1) \mathcal{O}_{-\beta}(z_2) {\mathcal E}(z_3) {\mathcal E}(z_4)} \nonumber \\
    & = \Braket{\mathcal{O}_{\beta}(z_1) \mathcal{O}_{-\beta}(z_2)} \Big[ \hat{\alpha}^{z_3,z_4} - (1-\cos\beta) \hat{\alpha}^{z_3,z_4}_{z_1|z_2} + \lambda (1-\cos\beta)^2 \hat{\alpha}^{z_3}_{z_1|z_2} \hat{\alpha}^{z_4}_{z_1|z_2} \Big],
\end{align}
where
\begin{equation} \label{def:alphahat3412}
    \hat{\alpha}^{z_3,z_4}_{z_1|z_2} := \hat{c}^2 \, \alpha^{z_3,z_4}_{z_1|z_2}
\end{equation}
with
\begin{equation} \label{def:alpha3412}
    \alpha^{z_3,z_4}_{z_1|z_2} \equiv \alpha^{z_3,z_4}_{z_2|z_1} := \lim_{\varepsilon \to 0} \varepsilon^{-4/3} \mu^{\text{loop}}(\ell \cap B_{\varepsilon}(z_j) \neq \emptyset \text{ for } j=3,4; \ell \text{ separates } z_1,z_2).
\end{equation}
The existence of the limit in \eqref{def:alpha3412} follows from the proof of Lemma \ref{lemma:conf-cov-mu}.

We note that $\alpha^{z_3,z_4}_{z_1|z_2} \equiv \alpha^{z_3,z_4}_{z_1|z_2;D}$ depends on the domain $D$. When $D={\mathbb C}$ we can determine $\alpha^{z_3,z_4}_{z_1|z_2}$ in terms of a quantity $Z_{\text{twist}}$, whose origin and meaning are explained in the next paragraph, and which was computed in \cite{Simmons_2009}. Using $Z_{\text{twist}}$, the weight can be written as
\begin{equation} \label{eq:alpha-Ztwist}
    \hat{\alpha}^{z_3,z_4}_{z_1|z_2;\mathbb C} = \frac{\hat{\alpha}^{z_3,z_4}_{\mathbb C}-Z_{\text{twist}}}{2},
\end{equation}
with
\begin{equation} \label{eq:alpha12bis}
    \hat{\alpha}^{z_3,z_4}_{\mathbb C} = \frac{1}{|z_3-z_4|^{4/3}},
\end{equation}
from \eqref{eq:EE-alpha}, \eqref{eq:canonical_edge}, and where
\begin{align} \label{ztwist}
    Z_{\text{twist}} = \left| \frac{z_{13} z_{24} }{ z_{34}^2 z_{23} z_{14} } \right|^{2/3} \left[
    \left| {}_2 F_1\left( -\frac{2}{3}, \frac{1}{3}; \frac{2}{3}, x \right)  \right|^2 - \frac{4\Gamma\left( \frac{2}{3} \right)^6}{\Gamma\left( \frac{4}{3} \right)^2 \Gamma\left( \frac{1}{3} \right)^4} |x|^{2/3} \left| {}_2 F_1\left( -\frac{1}{3}, \frac{2}{3}; \frac{4}{3}, x \right) \right|^2
    \right]
\end{align}
corresponds to Eq.\ (52) of \cite{Simmons_2009} with $x=\frac{z_{12}z_{34}}{z_{13}z_{24}}$.

In the language of \cite{Simmons_2009}, $Z_{\text{twist}}$ is the four-point function of a pair of ``2-leg'' operators $\phi_{0,1}$ with a pair of ``twist'' operators $\phi_{2,1}$\footnote{The subscripts label the positions of the operators in the Kac table.} in the $O(n)$ model in the limit $n \to 0$. The ``2-leg'' operator $\phi_{0,1}(z)$ forces a self-avoiding loop of the $O(n)$ model to go through $z$, while a pair of ``twist'' operators $\phi_{2,1}(z_1)\phi_{2,1}(z_2)$ acts like ${\mathcal O}_{\pi}(z_1){\mathcal O}_{-\pi}(z_2)$ in the sense that the weight of each loop that separates $z_1$ and $z_2$ is multiplied by $-1$.
Simmons and Cardy \cite{Simmons_2009} compute this four-point function for the $O(n)$ model for $-2<n<2$, which in the case of $n=0$ leads to \eqref{ztwist}. The $n=0$ case of the $O(n)$ model corresponds to a self-avoiding loop whose properties are described by $\mu^{\text{loop}}$, as we will now explain.

Strictly speaking, when $n=0$ all loops are suppressed, but the inclusion of a pair of 2-leg operators guarantees the presence of at least one loop.  Sending $n \to 0$ then singles out the ``one loop sector'' described by $\mu^{\text{loop}}$, since all other ``sectors'' produce a contribution of higher order in $n$ (see the discussion preceding Eq.\ (49) of \cite{Simmons_2009}).

Something analogous happens in the case of the four-point function \eqref{eq:OOEE}. As explained above, the pair of operators ${\mathcal O}_{\pi}(z_1){\mathcal O}_{-\pi}(z_2)$ acts like $\phi_{2,1}(z_1)\phi_{2,1}(z_2)$, while the presence of a pair of edge operators guarantees the existence of at least one loop. Since the loop soup can be thought of as a gas of loops in the grand canonical ensemble with fugacity $\lambda$, the four-point function can be written as a sum of contributions from various ``sectors'' characterized by the number of loops. Because of the normalization of the edge operator, which includes a factor of $\lambda^{-1/2}$, the contribution of the ``one loop sector'' is of order $O(1)$, while all other contributions are of order $O(\lambda)$, as one can clearly see from \eqref{eq:OOEE}. As a result, sending $\lambda \to 0$ in \eqref{eq:OOEE} singles out the ``one loop sector'' just like sending $n \to 0$ in the case of the $O(n)$ four-point function calculated by Simmons and Cardy \cite{Simmons_2009}. The two limits can be directly compared because all operators involved are canonically normalized. We can therefore conclude that
\begin{align}
\begin{split} \label{eq:Ztwist-explained}
    Z_{\text{twist}} & = \lim_{\lambda \to 0} \Braket{{\mathcal O}_{\pi}(z_1){\mathcal O}_{-\pi}(z_2){\mathcal E}(z_3){\mathcal E}(z_4)}_{\mathbb C} \\
    & = \hat{\alpha}^{z_3,z_4}_{\mathbb C} - 2\hat{\alpha}^{z_3,z_4}_{z_1|z_2;{\mathbb C}}.
\end{split}
\end{align}

In conclusion, equation \eqref{eq:OOEE} combined with \eqref{eq:alpha-Ztwist}-\eqref{ztwist} and \eqref{vvee_weights1} provides an explicit expression for the full-plane mixed four-point function $\Braket{\mathcal{O}_\beta(z_1) \mathcal{O}_{-\beta}(z_2) \mathcal{E}(z_3) \mathcal{E}(z_4)}_{\mathbb C}$.

\section{Higher-order and charged edge operators} \label{sec:higher_order&charged}

We will now extend the analysis of the edge operator $\mathcal{E}$ to all spin-zero operators that have non-zero fusion with the vertex operators. We will show that they have holomorphic and anti-holomorphic conformal dimensions
\begin{align}
    (\Delta(\beta) + k/3, \Delta(\beta) + k/3),
\end{align}
with $\Delta(\beta) = \frac{\lambda}{10}(1-\cos\beta)$, for any non-negative integer $k$. They correspond to the operators indicated on the diagonal of Figure \ref{VVVV_blocks}.
We will first define the operators with $\beta=0$ and dimensions $(k/3, k/3)$ for $k \geq 2$, which will be denoted $\mathcal{E}^{(k)}$ and will be called \emph{higher-order edge operators}. We will then see that the operators $\mathcal{E}_\beta^{(k)}$
with dimensions $(\Delta(\beta) + k/3, \Delta(\beta) + k/3)$ with $\beta \neq 0$ are a product of
 $  \mathcal{O}_\beta $ with a 
 modified version of $\mathcal{E}^{(k)}$. These will be called \emph{charged edge operators}.

\subsection{Higher-order edge operators} \label{sec:higher_order}

Searching for new primary operators, we are guided by their conformal dimensions. For the operators with dimensions $(k/3, k/3)$, it is natural to consider powers of edge operators. However, these are not well defined. Indeed, even if we keep both $\varepsilon$ and $\delta$ cutoffs, it is clear that $\big(E^{\delta}_{\varepsilon}(z)\big)^k$ is not the correct starting point because its mean is not zero.
A better choice, inspired by
\begin{align}
\begin{split} \label{E}
    E_{\varepsilon}^{(1);\delta}(z) := E_{\varepsilon}^{\delta}(z) & = N_{\varepsilon}^{\delta}(z) - \lambda\mu^{\text{loop}}(\text{diam}(\ell)>\delta, \ell \cap B_{\varepsilon}(z) \neq \emptyset) \\
    & = \left. \left(\frac{\partial}{\partial x} - \lambda\mu^{\text{loop}}(\text{diam}(\ell)>\delta, \ell \cap B_{\varepsilon}(z) \neq \emptyset)\right) x^{N^{\delta}_{\varepsilon}(z)} \right|_{x=1},
\end{split}
\end{align}
is given, for each integer $k \geq 2$, by
\begin{align}
\begin{split} \label{def:order-k-edge}
    E_{\varepsilon}^{(k);\delta}(z) & := \left. \Big(\frac{\partial}{\partial x} - \lambda\mu^{\text{loop}}(\text{diam}(\ell)>\delta, \ell \cap B_{\varepsilon}(z) \neq \emptyset)\Big)^k x^{N^{\delta}_{\varepsilon}(z)} \right|_{x=1} \\
    & = \sum_{j=0}^{k-1} (-1)^j \binom{k}{j}
    N_{\varepsilon}(z) \ldots (N_{\varepsilon}(z)-(k-j) +1)
    \left(\lambda\mu^{\text{loop}}(\text{diam}(\ell)>\delta, \ell \cap B_{\varepsilon}(z) \neq \emptyset) \right)^j \\
    & \quad + (-1)^k \left( \lambda\mu^{\text{loop}}(\text{diam}(\ell)>\delta, \ell \cap B_{\varepsilon}(z) \neq \emptyset) \right)^k.
\end{split}
\end{align}
This definition is valid in any domain $D$.  Since $N^{\delta}_{\varepsilon}(z)=n^{\varepsilon}_z({\mathcal L}^{\delta})$ (see Section \ref{sec:starmeasure} above \eqref{eq:E^*} and Appendix \ref{Appendix}) is a Poisson random variable with parameter $\lambda\mu^{\text{loop}}(\text{diam}(\ell)>\delta, \ell \cap B_{\varepsilon}(z) \neq \emptyset)$, we have that
\begin{align}
\Braket{N^{\delta}_{\varepsilon}(z) (N^{\delta}_{\varepsilon}(z)-1)\dots (N^{\delta}_{\varepsilon}(z)-(k-j) +1)} = \left( \lambda\mu^{\text{loop}}(\text{diam}(\ell)>\delta, \ell \cap B_{\varepsilon}(z) \neq \emptyset) \right)^{k-j},
\end{align}
which implies that $\Braket{E_{\varepsilon}^{(k);\delta}(z)}_\mathbb{C} = 0$ for every $\delta>0$.

With this notation, for each $k \geq 1$, we formally define the \emph{order $k$ edge operator} 
\begin{equation} \label{def:order_k_edge}
{\mathcal E}^{(k)}(z) := \frac{\hat{c}^k}{\sqrt{k!} \lambda^{k/2}} \lim_{\delta,\varepsilon \to 0} \vartheta_{\varepsilon}^{-k} E^{(k);\delta}_{\varepsilon}(z).
\end{equation}
As we will see at the end of this section, the constant in front of the limit is chosen in such a way that ${\mathcal E}^{(k)}$ is canonically normalized, i.e.,
\begin{equation}
    \Braket{\mathcal{E}^{(k)}(z_1)\mathcal{E}^{(k)}(z_2)}_{\mathbb C} = |z_1-z_2|^{-4k/3}.
\end{equation}
For $k=1$, we recover the edge operator, i.e., ${\mathcal E}^{(1)}\equiv{\mathcal E}$.

Definition \eqref{def:order_k_edge} is formal in the sense that ${\mathcal E}^{(k)}(z)$ is only well defined within $n$-point correlation functions. In order to show that ${\mathcal E}^{(k)}$ has well-defined $n$-point functions, we start with an intermediate result, for which we need the following notation. Given a collection of points $z_1, \ldots, z_n$ and a vector $\mathbf{k}=(k_1, \ldots, k_n)$, $k_j \in \mathbb{N}$, we denote by $\mathcal{M} \equiv \mathcal{M}(z_1,\ldots,z_n;k_1,\ldots,k_n)$ the collection of all multisets\footnote{A multiset is a set whose elements have multiplicity $\geq 1$.} $M$ such that
\begin{enumerate}
    \item[(1)] the elements $S$ of $M$ are subsets of $\{ z_1, \ldots, z_n \}$ with $|S|>1$,
    \item[(2)] the multiplicities $m_M(S)$ are such that $\sum_{S \in M} m_M(S) \mathbf{I}(z_j \in S) = k_j$ for each $j=1,\ldots,n$ and each $M \in \mathcal{M}$. 
\end{enumerate}
Condition (2) on the multiplicities essentially says that each point $z_j$ has multiplicity exactly $k_j$ in each multiset $M$. Note that $\mathcal{M}$ can be empty since conditions (1) and (2) cannot necessarily be satisfied simultaneously for generic choices of the vector $\mathbf{k}$.

For a set $S$, let $I_S$ denote the set of indices such that $j \in I_S$ if and only if $z_j \in S$. Then we have the following lemma, proved in the appendix.
\begin{lemma} \label{lemma:correlations-higher-cutoff}
For any $n \geq 2$ and $\delta,\varepsilon>0$, for any collection of points $z_1, \ldots, z_n$ at distance grater than $2\varepsilon$ from each other, with the notation introduced above, we have that
\begin{align}
\begin{split} \label{eq:correlations-charged-cutoff}
    & \Braket{ \prod_{j=1}^n E_{\varepsilon}^{(k_j)}(z_j) } := \lim_{\delta \to 0} \Braket{ \prod_{j=1}^n E_{\varepsilon}^{(k_j);\delta}(z_j) } \\
    & \quad = \left(\prod_{j=1}^n k_j!\right) \sum_{M \in {\mathcal M}} \prod_{S \in M} \frac{1}{m_M(S)!} \Big( \lambda \mu^{\text{loop}}(\ell \cap B_{\varepsilon}(z_j) \neq \emptyset \; \forall z_j \in S)\Big)^{{m_{M}(S)}} \mathbf{I}(\mathcal{M} \neq \emptyset),
\end{split}
\end{align}
where $\mathbf{I}(\mathcal{M} \neq \emptyset)$ denotes the indicator function of the event that $\mathcal{M}$ is not empty.
\end{lemma}

The next theorem shows that it is also possible to remove the $\varepsilon$ cutoff and demonstrates that the operators ${\mathcal E}^{(k)}$ are primaries with dimensions $(k/3,k/3)$ for all non-negative integer $k$.
\begin{theorem} \label{thm:correlations-higher_order-edge}
Let $D \subseteq{\mathbb C}$ be either the complex plane $\mathbb C$ or the upper-half plane $\mathbb H$ or any domain conformally equivalent to $\mathbb H$. With the notation of the previous lemma, for any collection of distinct points $z_1, \ldots, z_n \in D$ with $n \geq 2$ and any vector ${\bf k}=(k_1, \ldots, k_n)$ with $k_j \in \mathbb{N}$ such that $\mathcal M$ is not empty, we have that
\begin{align}
\begin{split} \label{eq:existence-higher-order}
& \mathcal{G}_D(z_1, \ldots, z_n;k_1, \ldots, k_n) := \lim_{\varepsilon \to 0} \vartheta_{\varepsilon}^{-\sum_{j=1}^n k_j} \Braket{ E^{(k_1)}_{\varepsilon}(z_1) \ldots E^{(k_n)}_{\varepsilon}(z_n) }_D \\
& \qquad \qquad = \left(\prod_{j=1}^n k_j! \right) \sum_{M \in {\mathcal M}} \prod_{S \in M} \frac{1}{m_M(S)!}  \big(\lambda\alpha^S \big)^{{m_{M}(S)}}.
\end{split}
\end{align}
Moreover, $\mathcal{G}_D(z_1, \ldots, z_n;k_1, \ldots, k_n)$ is conformally invariant in the sense that, if $D'$ is a domain conformally equivalent to $D$ and $f:D \to D'$ is a conformal map, then
\begin{align} \label{eq:conf-cov-higher}
    & \mathcal{G}_{D'}(f(z_1), \ldots, f(z_n);k_1, \ldots, k_n) \nonumber \\
    & \qquad = \left( \prod_{j=1}^n |f'(z_j)|^{-2k_j/3} \right) \mathcal{G}_D(z_1, \ldots, z_n;k_1, \ldots, k_n).
\end{align}
\end{theorem}

\noindent{\bf Proof.} From the expression for the $n$-point function in Lemma \ref{lemma:correlations-higher-cutoff}, using the fact that $\sum_{S \in M} m_M(S) \mathbf{I}(z_j \in S) = k_j,$ for each $j=1,\ldots,n$ and each $M \in {\mathcal M}$, we see that
\begin{align}
\begin{split}
    & \lim_{\varepsilon \to 0} \vartheta_{\varepsilon}^{-\sum_{j=1}^n k_j} \Braket{ \prod_{j=1}^n E_{\beta_j;\varepsilon}^{(k_j)}(z_j) } \\
    & \quad = \Big(\prod_{j=1}^n k_j!\Big) \sum_{M \in {\mathcal M}} \prod_{S \in M}\frac{1}{m_M(S)!}  \Big( \lambda \lim_{\varepsilon \to 0} \vartheta_{\varepsilon}^{-|S|} \mu^{\text{loop}}(\ell \cap B_{\varepsilon}(z_j) \neq \emptyset \; \forall z_j \in S)\Big)^{{m_{M}(S)}} \\
    & \quad = \Big(\prod_{j=1}^n k_j! \Big) \sum_{M \in {\mathcal M}} \prod_{S \in M} \frac{1}{m_M(S)!}  \big(\lambda\alpha^S \big)^{{m_{M}(S)}},
\end{split}
\end{align}
where the last equality follows from Lemma \ref{lemma:conf-cov-mu}.
Equation \eqref{eq:conf-cov-higher} now follows immediately from the last expression and Lemma \ref{lemma:conf-cov-mu}. \qed

Using \eqref{eq:existence-higher-order} and the definition of order $k$ edge operator \eqref{def:order_k_edge}, we can now write the correlation of $n$ higher-order edge operators as
\begin{align}
\begin{split}
    & \Braket{ \mathcal{E}^{(k_1)}(z_1) \ldots \mathcal{E}^{(k_n)}(z_n) }_{D} = \prod_{j=1}^n \frac{\hat{c}^{k_j}}{k_j!\lambda^{k_j/2}} \mathcal{G}_D(z_1, \ldots, z_n;k_1, \ldots, k_n) \\
    & \qquad \qquad = \left( \prod_{j=1}^n \lambda^{-k_j/2} \right) \sum_{M \in {\mathcal M}} \prod_{S \in M} \frac{1}{m_M(S)!}  \big(\lambda\hat{\alpha}^S \big)^{{m_{M}(S)}}. \label{eq:n-point-func-higher}
\end{split}
\end{align}
In view of \eqref{eq:conf-cov-higher}, these $n$-point functions are manifestly conformally covariant, showing that the higher-order edge operators are conformal primaries.

If $n=2$ and $k_1=k_2=k$, it is easy to see that the set $\mathcal{M}$ contains a single multiset with only one element $S=\{z_1,z_2\}$ with multiplicity $k$. Therefore,
specializing \eqref{eq:n-point-func-higher} to this case with $D=\mathbb{C}$ gives
\begin{align}
\begin{split}
    \Braket{ \mathcal{E}^{(k)}(z_1) \mathcal{E}^{(k)}(z_2) }_{\mathbb C} = \big( \hat\alpha^{z_1,z_2} \big)^k = \left( \Braket{ \mathcal{E}(z_1) \mathcal{E}(z_2) }_{\mathbb C} \right)^k = |z_1-z_2|^{-4k/3},
\end{split}
\end{align}
which shows that $\mathcal{E}^{(k)}$ is canonically normalized.

\subsection{Charged edge operators} \label{sec:charged}

We now apply a ``twist'' to the (higher-order) edge operators and introduce a new set of operators. A \emph{charged edge operator} is essentially an edge operator ``seen from'' the perspective of a measure $\mu^*_{z;\beta} \equiv \mu^*_{z}$ defined by
\begin{equation} \label{mu-star2}
\mu^*_{z}(\ell) := \left\{
\begin{array}{lll}
    \mu^{\text{loop}}(\ell) & \mbox{if $z \notin \bar\ell$} \\
    e^{i\beta\sigma_{\ell}} \mu^{\text{loop}}(\ell) & \mbox{if $z \in \bar{\ell}$}
\end{array} \right.
\end{equation}
where $\sigma_{\ell} = \pm 1$ is a symmetric Boolean variable assigned to $\ell$.
This measure, which is similar to that introduced in Section \ref{sec:starmeasure}, assigns a phase $e^{i\beta\sigma_{\ell}}$ to each loop covering $z$.

We note that, when taking expectations, one sums over the two possible values of $\sigma_{\ell}$ with equal probability, so that loops $\ell$ that do not cover $z$ get weight $\mu^{\text{loop}}(\ell)$, while loops $\ell$ that cover $z$ get weight $\cos\beta \, \mu^{\text{loop}}(\ell)$.

With this in mind, for any $\beta \in [0,2\pi)$, the simplest charged edge operator with cutoffs $\delta,\varepsilon>0$, corresponding to the ``twisted'' or ``charged'' version of \eqref{E}, is defined as
\begin{align}
\begin{split}
& E_{\beta;\varepsilon}^{(1);\delta}(z) \equiv E^{\delta}_{\beta;\varepsilon}(z) \\
& \quad := V^{\delta}_{\beta}(z) \Big(N^{\delta}_{\varepsilon}(z) \\
& \qquad - \lambda \big(\mu^{\text{loop}}(\text{diam}(\ell)>\delta, \ell \cap B_{\varepsilon}(z) \neq \emptyset, z \notin \overline \ell) + \mu^{\text{loop}}(\ell \cap B_{\varepsilon}(z) \neq \emptyset, z \in \overline \ell) \cos\beta \big) \Big),
\end{split}
\end{align}
where
\begin{align}
    V^{\delta}_{\beta}(z) := \exp{\Big(i \beta \sum_{\stackrel{\ell \in \mathcal{L}^{\delta} }{z \in \bar\ell}} \sigma_{\ell} \Big)},
\end{align}
the layering operator with cutoff $\delta>0$ introduced in \cite{Camia_2016}, induces a phase $e^{i \beta\sigma_{\ell}}$ for each loop $\ell$ such that $z \in \bar\ell$, and 
\begin{align}
    \lambda \big(\mu^{\text{loop}}(\text{diam}(\ell)>\delta, \ell \cap B_{\varepsilon}(z) \neq \emptyset, z \notin \overline \ell) + \mu^{\text{loop}}(\ell \cap B_{\varepsilon}(z) \neq \emptyset, z \in \overline \ell) \cos\beta \big)
\end{align}
is the expectation of $N^{\delta}_{\varepsilon}(z)$ under the measure $\mu^*_z$.

Generalizing this to any $k \in \mathbb{N}$, the ``twisted'' or ``charged'' version of \eqref{def:order-k-edge} is given by
\begin{align} \label{def:order-k-edge_charged}
\begin{split}
    & E^{(k);\delta}_{\beta;\varepsilon}(z) := V^{\delta}_{\beta}(z)\Bigg[\sum_{j=0}^{k-1} (-1)^j \binom{k}{j} N^{\delta}_{\varepsilon}(z) \ldots (N^{\delta}_{\varepsilon}(z)-(k-j) + 1) \\
    & \left(\lambda \big(\mu^{\text{loop}}(\text{diam}(\ell)>\delta, \ell \cap B_{\varepsilon}(z) \neq \emptyset, z \notin \overline \ell) + \mu^{\text{loop}}(\ell \cap B_{\varepsilon}(z) \neq \emptyset, z \in \overline \ell) \cos\beta \big) \right)^j \\
    & + (-1)^k \left( \lambda \big(\mu^{\text{loop}}(\text{diam}(\ell)>\delta, \ell \cap B_{\varepsilon}(z) \neq \emptyset, z \notin \overline \ell) + \mu^{\text{loop}}(\ell \cap B_{\varepsilon}(z) \neq \emptyset, z \in \overline \ell) \cos\beta \big) \right)^k \Bigg].
\end{split}
\end{align}

We now formally define the \emph{charged (order $k$) edge operator}
\begin{equation} \label{def:charged_edge_limit}
{\mathcal E}^{(k)}_{\beta}(z) := \lim_{\delta,
\varepsilon \to 0} (c'\delta)^{-2\Delta(\beta)} 
\frac{\hat{c}^k}{k!\lambda^{k/2}} \vartheta_{\varepsilon}^{-k} E^{(k);\delta}_{\beta;\varepsilon}(z),
\end{equation}
where $c'$ is a normalization constant needed to obtain the canonically normalized operator $\mathcal{O}_{\beta}$ from $V^{\delta}_{\beta}$, which depends on the domain (see \cite{Camia_2020}). For completeness, we also define $\mathcal{E}_{\beta}^{(0)} \equiv \mathcal{O}_{\beta}$.
Unlike their uncharged counterparts, the charged operators $\mathcal{E}_{\beta}^{(k)}$ are not canonically normalized for general $\beta \neq 0$.

As an example, we compute the two-point function of the simplest charged edge operators, with charge conservation. To that end, we write $E^{\delta}_{\beta;\varepsilon}(z)$ as
\begin{align}
\begin{split} \label{eq:chargededge}
& E^{\delta}_{\beta;\varepsilon}(z) = V^{\delta}_{\beta}(z) \Big(N^{\delta}_{\varepsilon}(z) - \lambda \mu^{\text{loop}}(\text{diam}(\ell)>\varepsilon, \ell \cap B_{\varepsilon}(z) \neq \emptyset) \\
& \qquad \qquad + (1-\cos\beta)\lambda\mu^{\text{loop}}(\text{diam}(\ell)>\delta, \ell \cap B_{\varepsilon}(z) \neq \emptyset, z \in \overline \ell)\Big) \\
& \qquad \quad = V^{\delta}_{\beta}(z) E^{\delta}_{\varepsilon}(z) + (1-\cos\beta)\lambda\mu^{\text{loop}}(\text{diam}(\ell)>\delta, \ell \cap B_{\varepsilon}(z) \neq \emptyset, z \in \overline \ell) V^{\delta}_{\beta}(z).
\end{split}
\end{align}

Using this expression and the method introduced in Section \ref{sec:starmeasure}, we have
\begin{align}
\begin{split}
& \Braket{E^{\delta}_{\beta;\varepsilon}(z_1) E^{\delta}_{-\beta;\varepsilon}(z_2)} = \Braket{V^{\delta}_{\beta}(z_1) V^{\delta}_{-\beta}(z_2) E^{\delta}_{\varepsilon}(z_1) E^{\delta}_{\varepsilon}(z_2)} \\
& \qquad + (1-\cos\beta)\lambda\mu^{\text{loop}}(\text{diam}(\ell)>\delta, \ell \cap B_{\varepsilon}(z_2) \neq \emptyset, z_2 \in \overline \ell) \Braket{V^{\delta}_{\beta}(z_1) E^{\delta}_{\varepsilon}(z_1) V^{\delta}_{-\beta}(z_2)} \\
& \qquad + (1-\cos\beta)\lambda\mu^{\text{loop}}(\text{diam}(\ell)>\delta, \ell \cap B_{\varepsilon}(z_1) \neq \emptyset, z_1 \in \overline \ell) \Braket{V^{\delta}_{-\beta}(z_2) E^{\delta}_{\varepsilon}(z_2) V^{\delta}_{\beta}(z_1)} \\
& \qquad + (1-\cos\beta)^2\lambda^2 \mu^{\text{loop}}(\text{diam}(\ell)>\delta, \ell \cap B_{\varepsilon}(z_1) \neq \emptyset, z_1 \in \overline \ell) \\
& \qquad \qquad \mu^{\text{loop}}(\text{diam}(\ell)>\delta, \ell \cap B_{\varepsilon}(z_2) \neq \emptyset, z_2 \in \overline \ell) \Braket{V^{\delta}_{\beta}(z_1) V^{\delta}_{-\beta}(z_2)} \\
& \qquad = \Braket{V^{\delta}_{\beta}(z_1) V^{\delta}_{-\beta}(z_2)} \Big[ \Braket{E^{\delta}_{\varepsilon}(z_1) E^{\delta}_{\varepsilon}(z_2)}^*_{z_1,z_2} \\
& \qquad + (1-\cos\beta)\lambda\mu^{\text{loop}}(\text{diam}(\ell)>\delta, \ell \cap B_{\varepsilon}(z_2) \neq \emptyset, z_2 \in \overline \ell) \Braket{E^{\delta}_{\varepsilon}(z_1)}^*_{z_1,z_2} \\
& \qquad + (1-\cos\beta)\lambda\mu^{\text{loop}}(\text{diam}(\ell)>\delta, \ell \cap B_{\varepsilon}(z_1) \neq \emptyset, z_1 \in \overline \ell) \Braket{E^{\delta}_{\varepsilon}(z_2)}^*_{z_1,z_2} \\
& \qquad + (1-\cos\beta)^2\lambda^2 \mu^{\text{loop}}(\text{diam}(\ell)>\delta, \ell \cap B_{\varepsilon}(z_1) \neq \emptyset, z_1 \in \overline \ell) \\
& \qquad \qquad \mu^{\text{loop}}(\text{diam}(\ell)>\delta, \ell \cap B_{\varepsilon}(z_2) \neq \emptyset, z_2 \in \overline \ell) \Big].
\end{split}
\end{align}
After identifying $z_3$ with $z_1$ and $z_4$ with $z_2$, we can use \eqref{eq:E^*} and \eqref{eq:EE^*} to simplify the above expression. A simple calculation shows that, for for any $\delta<|z_1-z_2|$,
\begin{align}
    \begin{split}
        & \Braket{E^{\delta}_{\beta;\varepsilon}(z_1) E^{\delta}_{-\beta;\varepsilon}(z_2)} = \Braket{V^{\delta}_{\beta}(z_1) V^{\delta}_{-\beta}(z_2)} \Big[ \lambda\mu^{\text{loop}}(\ell \cap B_{\varepsilon}(z_j) \neq \emptyset, j=1,2) \\
        & \quad - (1-\cos\beta)\lambda\mu^{\text{loop}}(\ell \cap B_{\varepsilon}(z_j) \neq \emptyset, j=1,2; \ell \text{ separates } z_1,z_2) \\
        & \quad + \lambda^2 (1-\cos\beta)^2 \mu^{\text{loop}}(\ell \cap B_{\varepsilon}(z_1) \neq \emptyset, z_2 \in \bar\ell, z_1 \notin \bar\ell) \\
        & \qquad \mu^{\text{loop}}(\text{diam}(\ell)>\delta, \ell \cap B_{\varepsilon}(z_2) \neq \emptyset, z_1 \in \bar\ell, z_2 \notin \bar\ell) \Big].
    \end{split}
\end{align}
Using definition \eqref{def:charged_edge_limit}, we obtain
\begin{align} \label{eq:charged-2-point-func}
    \begin{split}
        & \Braket{\mathcal{E}_{\beta}(z_1) \mathcal{E}_{-\beta}(z_2)} = \lim_{\delta \to 0} (\hat{c}'\delta)^{-4\Delta(\beta)} \Braket{V^{\delta}_{\beta}(z_1) V^{\delta}_{-\beta}(z_2)} \\
        & \quad \hat{c}^2 \lim_{\varepsilon \to 0} \vartheta_{\varepsilon}^{-2} \Big[ \mu^{\text{loop}}(\ell \cap B_{\varepsilon}(z_j) \neq \emptyset, j=1,2) \\
        & \qquad - (1-\cos\beta) \mu^{\text{loop}}(\ell \cap B_{\varepsilon}(z_j) \neq \emptyset, j=1,2; \ell \text{ separates } z_1,z_2) \\
        & \qquad + \lambda (1-\cos\beta)^2 \mu^{\text{loop}}(\ell \cap B_{\varepsilon}(z_1) \neq \emptyset, z_2 \in \bar\ell, z_1 \notin \bar\ell) \\
        & \qquad \quad \mu^{\text{loop}}(\ell \cap B_{\varepsilon}(z_2) \neq \emptyset, z_1 \in \bar\ell, z_2 \notin \bar\ell) \Big] \\
        & = \Braket{\mathcal{O}_{\beta}(z_1) \mathcal{O}_{-\beta}(z_2)} \Big[ \hat\alpha^{z_1,z_2} - (1-\cos\beta) \hat\alpha^{z_1,z_2}_{z_1|z_2} \\
        & \qquad + \lambda (1-\cos\beta)^2 \hat{c}^2 \lim_{\varepsilon \to 0} \varepsilon^{-4/3} \mu^{\text{loop}}(\ell \cap B_{\varepsilon}(z_1) \neq \emptyset, z_2 \in \bar\ell, z_1 \notin \bar\ell) \\
        & \qquad \quad \mu^{\text{loop}}(\ell \cap B_{\varepsilon}(z_2) \neq \emptyset, z_1 \in \bar\ell, z_2 \notin \bar\ell) \Big].
    \end{split}
\end{align}

At this point, we should note that unfortunately the existence of the limits
\begin{align} \label{two-limits}
    \begin{split}
        & \alpha^{z_1,z_2}_{z_1|z_2} = \lim_{\varepsilon \to 0} \vartheta_{\varepsilon}^{-2} \mu^{\text{loop}}(\ell \cap B_{\varepsilon}(z_j) \neq \emptyset, j=1,2; \ell \text{ separates } z_1,z_2), \\
        & \lim_{\varepsilon \to 0} \vartheta_{\varepsilon}^{-1} \mu^{\text{loop}}(\ell \cap B_{\varepsilon}(z_j) \neq \emptyset, z_k \in \bar\ell, z_j \notin \bar\ell)
    \end{split}
\end{align}
does not follow from Lemma \ref{lemma:conf-cov-mu}.
It is, however, reasonable to assume that they exist. Indeed, in the case of the first limit, observing that 
\begin{align}
    \lim_{\substack{z_3 \to z_1 \\ z_4 \to z_2}} Z_\text{twist} = 0
\end{align}
and using \eqref{eq:alpha-Ztwist} suggests that, in the full plane,
\begin{align} \label{eq:alpha-limit}
    \hat\alpha^{z_1,z_2}_{z_1|z_2;\mathbb{C}} = \frac{1}{2} \hat\alpha^{z_1,z_2}_{\mathbb C}.
\end{align}
The second limit in \eqref{two-limits} should also exist; moreover, if
\begin{align}
    \begin{split}
        \hat\alpha^{z_j}_{\mathbb{C}}(z_k;z_j) := \hat{c} \lim_{\varepsilon \to 0} \vartheta_{\varepsilon}^{-1} \mu_{\mathbb{C}}^{\text{loop}}(\ell \cap B_{\varepsilon}(z_j) \neq \emptyset, z_k \in \bar\ell, z_j \notin \bar\ell)
    \end{split}
\end{align}
does exist, arguments like those used in the second part of the proof of Lemma \ref{lemma:conf-cov-mu} imply that, for any $s>0$, $\hat\alpha^{sz}_{\mathbb{C}}(0;z) = s^{-2/3} \hat\alpha^{z}_{\mathbb{C}}(0;z)$. Since $\hat\alpha^{z_j}_{\mathbb{C}}(z_k;z_j)$ only depends on $|z_j-z_k|$, this would in turn imply that $\hat\alpha^{z_j}_{\mathbb{C}}(z_k;z_j)$ must take the form $\text{const } |z_j-z_k|^{-2/3}$.

If the considerations above are correct, then it follows from \eqref{eq:charged-2-point-func} that $\Braket{\mathcal{E}_{\beta}(z_1) \mathcal{E}_{-\beta}(z_2)}_{\mathbb{C}}$ behaves like the correlation function between two conformal primaries of scaling dimension $\Delta(\beta)+1/3$, as desired.
Indeed, we conjecture that, similarly to \eqref{eq:alpha-limit}, $\hat\alpha^{z_j}_{\mathbb{C}}(z_k;z_j) = \frac{1}{2}\hat\alpha^{z_j}_{z_k;\mathbb{C}}$, which would lead to
\begin{align}
\begin{split}
        \Braket{\mathcal{E}_{\beta}(z_1) \mathcal{E}_{-\beta}(z_2)}_{\mathbb{C}} &= \Braket{\mathcal{O}_{\beta}(z_1) \mathcal{O}_{-\beta}(z_2)}_{\mathbb{C}} \left( \frac{1}{2} (1+\cos\beta) \hat\alpha^{z_1,z_2}_{\mathbb C} + \frac{\lambda}{4} \hat\alpha^{z_1}_{z_2;\mathbb C} \hat\alpha^{z_2}_{z_1;\mathbb C} \right) \\
        &\sim |z_1-z_2|^{-4\Delta(\beta)-4/3},
\end{split}
\end{align}
where the existence and the scaling behavior of
\begin{align}
    \begin{split}
        \hat\alpha^{z_j}_{z_k;\mathbb{C}} := \hat{c} \lim_{\varepsilon \to 0} \vartheta_{\varepsilon}^{-1} \mu_{\mathbb{C}}^{\text{loop}}(\ell \cap B_{\varepsilon}(z_j) \neq \emptyset, z_k \in \bar\ell)
    \end{split}
\end{align}
follow from the proof of Lemma \ref{lemma:conf-cov-mu}.

\section{The primary operator spectrum} \label{sec:spectrum}

The four-point function of a conformal field theory contains information about the three-point function coefficients, as well as the spectrum of primary operators. In the following two sections, we perform the Virasoro conformal block expansion of the new four-point function \eqref{eq:OOEE} in the full plane, and derive the three-point coefficient involving three edge operators through the OPE of the edge operator as an illustration of the conformal block expansion.

\subsection{Virasoro conformal blocks} \label{sec:conf-blocks}

By a global conformal transformation, one can always map three of the four points of a four-point function $\Braket{\mathcal{A}_1(z_1) \mathcal{A}_2(z_2) \mathcal{A}_3(z_3) \mathcal{A}_4(z_4)}_\mathbb{C}$ to fixed values, where $\mathcal{A}_j(z_j)$ here denotes a generic primary operator evaluated at $z_j$. The remaining dependence is only on the cross-ratio $x=\frac{z_{12}z_{34}}{z_{13}z_{24}}$ and its complex conjugate $\bar{x}$, which are invariant under global conformal transformations.
The following discussion parallels Section 6 of \cite{Camia_2020}.
Following the notation of Section $6.6.4$ of \cite{DiFrancesco:639405}, it is customary to set $z_1 = \infty, ~z_2 = 1, ~z_3 = x$ and $z_4 = 0$. The resulting function
\begin{align}
    G^{21}_{34}(x) := \lim_{z_1 \to \infty} z_1^{2\Delta_1} \bar{z}_1^{2\bar{\Delta}_1} \Braket{\mathcal{A}_1(z_1) \mathcal{A}_2(1) \mathcal{A}_3(x) \mathcal{A}_4(0)}_\mathbb{C}
\end{align}
can be decomposed into Virasoro conformal blocks according to
\begin{align} \label{virasoroblocks}
    G^{21}_{34}(x) = \sum_\mathcal{P} C_{34}^\mathcal{P} C_{12}^\mathcal{P} \mathcal{F}_{34}^{21}(\mathcal{P}|x) \bar{\mathcal{F}}_{34}^{21}(\mathcal{P}|\bar{x}).
\end{align}

The sum over $\mathcal{P}$ runs over all primary operators in the theory, and the $C_{lj}^\mathcal{P}$ are the three-point function coefficients of the operators labeled by $l,j, \mathcal{P}$, that is,
\begin{align}
\begin{split}
    \Braket{\mathcal{A}_l(z_1) \mathcal{A}_j(z_2) \mathcal{P}(z_3)}_\mathbb{C} = C_{lj}^\mathcal{P} &z_{12}^{-(\Delta_l+\Delta_j-\Delta_{\mathcal{P}})} z_{13}^{-(\Delta_l+\Delta_{\mathcal{P}}-\Delta_j)} z_{23}^{-(\Delta_j+\Delta_{\mathcal{P}}-\Delta_l)} \\
    &\bar{z}_{12}^{-(\bar{\Delta}_l+\bar{\Delta}_j-\bar{\Delta}_{\mathcal{P}})} \bar{z}_{13}^{-(\bar{\Delta}_l+\bar{\Delta}_{\mathcal{P}}-\bar{\Delta}_j)} \bar{z}_{23}^{-(\bar{\Delta}_j+\bar{\Delta}_{\mathcal{P}}-\bar{\Delta}_l)},
\end{split}
\end{align}
where $\Delta_j,\bar{\Delta}_j$ are the scaling dimensions of the corresponding fields.

The functions $\mathcal{F},\bar{\mathcal{F}}$ are called Virasoro conformal blocks and are fixed by conformal invariance. Each conformal block can be written as a power series
\begin{align} \label{fblock}
    \mathcal{F}^{21}_{34}(\mathcal{P}|x) = x^{\Delta_\mathcal{P} - \Delta_3 - \Delta_4} \sum_{K=0}^\infty \mathcal{F}_K x^K,
\end{align}
where coefficients $\mathcal{F}_K$ can be fully determined by the central charge $c$, and the conformal dimensions $\Delta_j, \Delta_\mathcal{P}$ of the five operators involved. $\bar{\mathcal{F}}$ is determined analogously.

\begin{figure}[t]
    \centering
    \begin{subfigure}[b]{0.4\textwidth}
        \includegraphics[width=\textwidth]{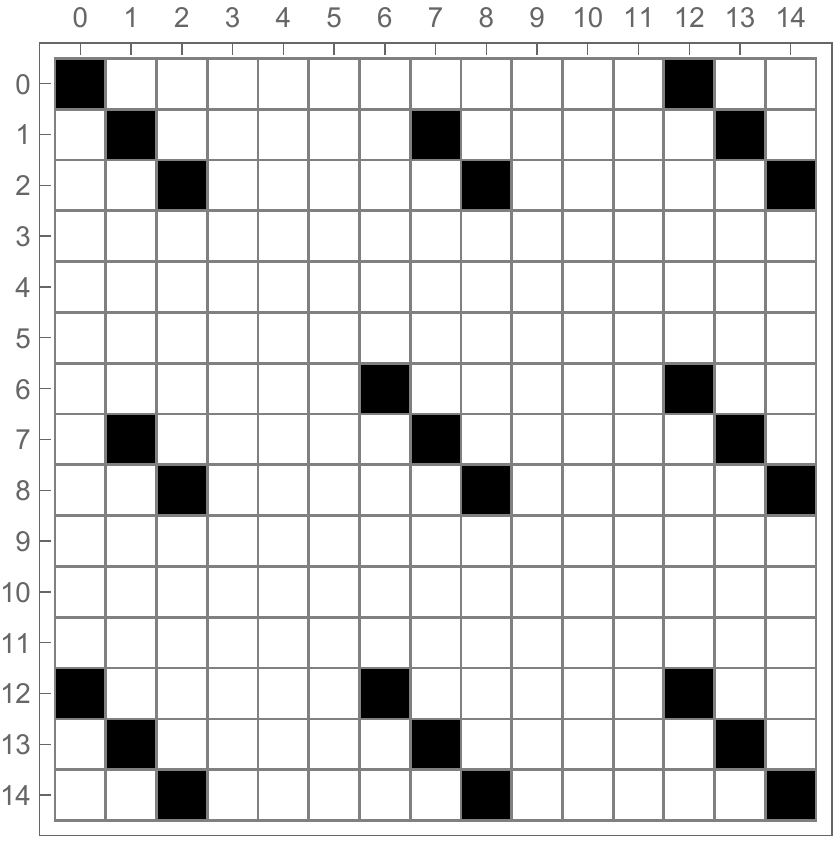}
        \caption{Non-zero $C_{\mathcal{E}\mathcal{E}}^{(p,p')}$}
        \label{VVEE_blocks}
    \end{subfigure} ~
    \begin{subfigure}[b]{0.4\textwidth}
        \includegraphics[width=\textwidth]{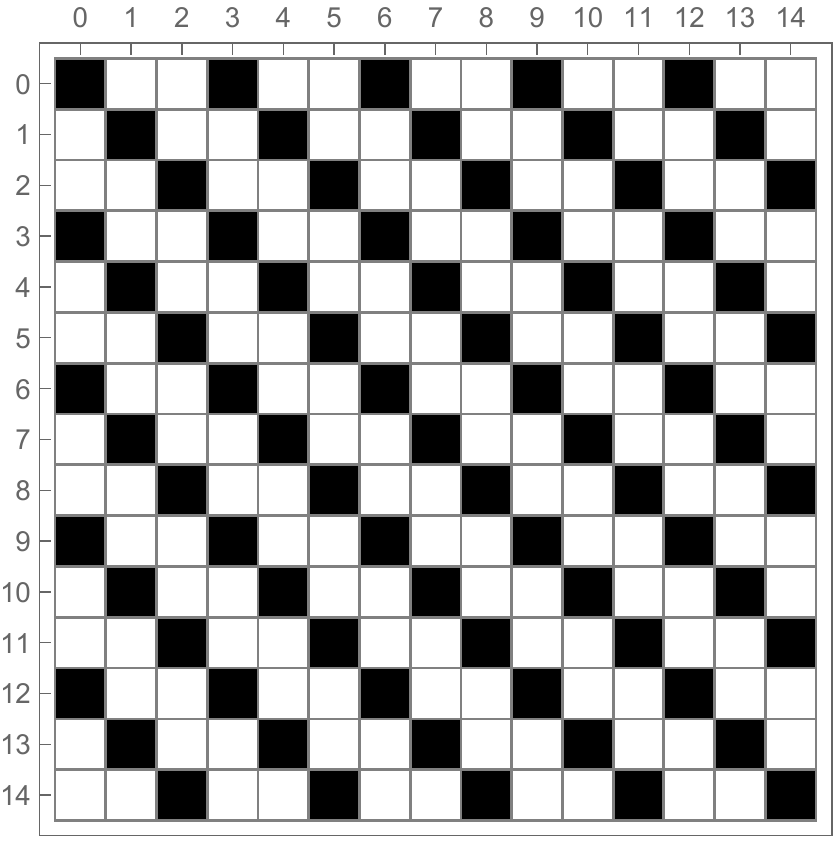}
        \caption{Non-zero $C_{\mathcal{O}_\beta \mathcal{O}_{-\beta}}^{(p,p')}$}
        \label{VVVV_blocks}
    \end{subfigure}
    \caption{The non-zero three-point function coefficients are shown. Rows and columns label $(p,p')$. Left: between two edge operators. Right: between two vertex operators.
    }
    \label{checkerboard}
\end{figure}

In the case of \eqref{eq:OOEE}, we obtain
\begin{align}
\begin{split} \label{defG1}
    & G^{21}_{34}(x) = \lim_{z_1 \rightarrow \infty} |z_1|^{4 \Delta(\beta)}
    \Braket{ \mathcal{O}_{\beta}(z_1) \mathcal{O}_{-\beta}(1) \mathcal{E}(x) \mathcal{E}(0)}_\mathbb{C} \\
    & = \lambda \frac{4 \cdot 2^{1/3} \pi^2}{5 \sqrt{3} \Gamma\left(\frac{1}{6}\right)^2 \Gamma\left(\frac{4}{3}\right)^2} \frac{ (1-\cos\beta)^2}{|1-x|^{2/3}} + \frac{1 + \cos\beta}{2|x|^{4/3}} \\
    & \quad + \frac{1-\cos\beta}{2|x|^{4/3} |1-x|^{2/3}}
    \left[ \left| {}_2 F_1\left( -\frac{2}{3}, \frac{1}{3}; \frac{2}{3}; x \right) \right|^2 
    - \frac{4 \Gamma\left( \frac{2}{3} \right)^6}{\Gamma\left( \frac{4}{3}\right)^2 \Gamma\left( \frac{1}{3}\right)^4} |x|^{2/3} \left| {}_2 F_1\left( -\frac{1}{3}, \frac{2}{3}; \frac{4}{3}; x \right) \right|^2
    \right].
\end{split}
\end{align}
The expansion around $x=\bar{x}=0$ allows us to obtain information about the primary operator spectrum and fusion rules of the operators that appear  in both the  $\mathcal{O}_\beta \times \mathcal{O}_{-\beta}$ and $\mathcal{E} \times \mathcal{E}$ expansions. The hypergeometric functions appearing above are regular around $x=0$. 
The expansion of \eqref{defG1} around zero can thus be written
\begin{align}
    G^{21}_{34}(x) = |x|^{-4/3} \sum_{m,n=0}^\infty a_{m,n}x^{m/3}\bar{x}^{n/3}.
\end{align}
Using \eqref{fblock}, this expansion is of the form $|x|^{- 4 \Delta_\mathcal{E}} x^{\Delta_\mathcal{P}+k} \bar{x}^{\bar{\Delta}_\mathcal{P}+\bar{k}}$, where $k,\bar{k}$ are non-negative integers. Since $\Delta_\mathcal{E} = 1/3$ we see that $\Delta_\mathcal{P},\bar{\Delta}_\mathcal{P}$ can only be multiples of 1/3.
This must be equal to \eqref{virasoroblocks}, which can now be written as
\begin{align} \label{gexpansion}
    G^{21}_{34}(x) = |x|^{-4/3} \sum_{\substack{p,p',\\m,n=0}}^\infty C_{\mathcal{E}\mathcal{E}}^{(p,p')} C_{\mathcal{O}_\beta \mathcal{O}_{-\beta}}^{(p,p')} \mathcal{F}_m^{(p)} \mathcal{F}_n^{(p')} x^{m/3}\bar{x}^{n/3}.
\end{align}
By comparing the last two equations, we determine the products of three-point function coefficients at any desired order. Together with the three-point coefficients determined in \cite{Camia_2020}, using \cite{headrick}, we can uniquely determine the coefficients involving edge operators which also fuse onto vertex operators.
Figure \ref{checkerboard} shows the non-zero three-point coefficients $C_{\mathcal{E}\mathcal{E}}^{(p,p')}$ which appear in the Virasoro block expansion. The operators appearing in Figure \ref{VVEE_blocks} are a subset of those in Figure \ref{VVVV_blocks}, and only the operators which fuse onto both sets of operators can be discovered from \eqref{defG1}.

The correct normalization of our operators and four-point function is ensured by
\begin{align}
\begin{split}
    C_{\mathcal{E}\mathcal{E}}^{(0,0)} &\equiv C_{\mathcal{E}\mathcal{E}}^\mathbb{1} = 1 \\
    C_{\mathcal{O}_\beta \mathcal{O}_{-\beta}}^{(0,0)} &\equiv C_{\mathcal{O}_\beta \mathcal{O}_{-\beta}}^\mathbb{1} = 1.
\end{split}
\end{align}
Furthermore, we obtain the coefficients
\begin{align} \label{CEEE}
    C_{\mathcal{E}\mathcal{E}}^{(1,1)} &\equiv C_{\mathcal{E}\mathcal{E}}^{\mathcal{E}} = \frac{1}{\sqrt{\lambda}} \frac{4 \cdot 2^{1/6} \cdot 3^{1/4} \cdot \sqrt{5} \pi^{3/2} \Gamma\left(\frac{2}{3}\right)}{\Gamma\left(\frac{1}{6}\right)^3 \Gamma\left(\frac{7}{6}\right)} \\
    C_{\mathcal{E}\mathcal{E}}^{(2,2)} &\equiv C_{\mathcal{E}\mathcal{E}}^{\mathcal{E}^{(2)}} = \sqrt{2}.
\end{align}
The complexity of these coefficients grows rapidly for larger $(p,p')$.
The operator $\mathcal{E}^{(2)}$ can be identified with the higher order edge operator of conformal and anti-conformal dimensions $2/3$ defined in \eqref{def:order_k_edge}.

By rearranging the operators in the four-point function \eqref{defG1}, one can easily show that the resulting four-point functions are crossing-symmetric. In particular, by exchanging operators 2 and 4, one may obtain information about the OPE of $\mathcal{O}_\beta \times \mathcal{E}$. The expansion in the cross-ratio in this channel shows logarithmic terms, which indicate the existence of degenerate operators in a logarithmic CFT. The logarithmic properties of the related $O(n)$ model have been studied, for example, in \cite{Gorbenko_2020}. We do not investigate their relations to the BLS at this point.

Nevertheless, one can use $G_{32}^{41}(x) = G_{34}^{21}(1-x)$ to compute the fusion rules for $\mathcal{O}_\beta \times \mathcal{E}$, and in particular, the squares of three-point function coefficients $C_{\mathcal{O}_{\beta} \mathcal{E}}^\mathcal{P}$ of all primaries $\mathcal{P}$.
The expansion of $G_{34}^{21}(1-x)$ analogous to \eqref{gexpansion} allows us to obtain the following operators in the OPE
\begin{align} \label{eq:OEOPE}
    \mathcal{O}_\beta(z) \times \mathcal{E}(z') = 
    C_{\mathcal{O}_\beta \mathcal{E}}^{\mathcal{O}_\beta} |z-z'|^{-2/3} \mathcal{O}_\beta(z)
    + C_{\mathcal{O}_\beta \mathcal{E}}^{\mathcal{E}_\beta} \mathcal{E}_\beta(z) + \ldots,
\end{align}
where $C_{\mathcal{O}_\beta \mathcal{E}}^{\mathcal{O}_\beta} = C_{\mathcal{O}_\beta \mathcal{O}_{-\beta}}^{\mathcal{E}}$ and
\begin{align}
    \left( C_{\mathcal{O}_\beta \mathcal{E}}^{\mathcal{E}_\beta} \right)^2 = \frac{1 + \cos\beta}{2}.
\end{align}
The operator $\mathcal{E}_\beta$ is the $k=1$ case of the charged edge operators defined in (\ref{def:charged_edge_limit}), with conformal and anti-conformal dimension $\Delta(\beta) + 1/3$.

\subsection{The three-point function of the edge operator} \label{sec:3edges}

In this section, we show how to compute the three-point function coefficient $C^{(1,1)}_{{\mathcal E}{\mathcal E}} \equiv C^{\mathcal E}_{{\mathcal E}{\mathcal E}}$, which was derived in the previous section from the conformal block expansion, by applying the OPE of two edge operators. This computation is a special case of the general expansion \eqref{gexpansion}, and shows the inner workings of the general method.

Using the general expression for the three-point function of a conformal primary operator and \eqref{eq:EE}, we have
\begin{align}
\begin{split}
    \Braket{{\mathcal E}(z_1){\mathcal E}(z_2){\mathcal E}(z_3)}_{\mathbb C} & = C^{\mathcal E}_{{\mathcal E}{\mathcal E}} \, |z_{12}|^{-2/3} \, |z_{13}|^{-2/3} \, |z_{23}|^{-2/3} \\
    & = C^{\mathcal E}_{{\mathcal E}{\mathcal E}} \, |z_{12}|^{-4/3} \, |z_{23}|^{-2/3} \, \left( 1 + O\big(|z_{23}|\big) \right) \\
    & = C^{\mathcal E}_{{\mathcal E}{\mathcal E}} \, \Braket{{\mathcal E}(z_1){\mathcal E}(z_2)}_{\mathbb C} \, |z_{23}|^{-2/3} \, \left( 1 + O\big(|z_{23}|\big) \right) \\
    & = \Braket{{\mathcal E}(z_1) \Big[C^{\mathcal E}_{{\mathcal E}{\mathcal E}} \, |z_{23}|^{-2/3} \, {\mathcal E}(z_2) + O\big( |z_{23}|^{1/3} \big) \Big]}_{\mathbb C}.
\end{split}
\end{align}
Additionally, using \eqref{eq:OOEE} and \eqref{eq:alpha-Ztwist} we see that, for $\beta=\pi$,
\begin{align}
\begin{split} \label{eq:OOEEpi}
    & \Braket{\mathcal{O}_{\pi}(z_1) \mathcal{O}_{-\pi}(z_2) \mathcal{E}(z_3) \mathcal{E}(z_4)}_{\mathbb C} \\
    & \quad = |z_{12}|^{-4\lambda/5} Z_{\text{twist}} + 4 \lambda \, |z_{12}|^{-4\lambda/5} \, \hat{\alpha}^{z_3}_{z_1|z_2;{\mathbb C}} \hat{\alpha}^{z_4}_{z_1|z_2;{\mathbb C}}.
\end{split}
\end{align}
The second term on the right-hand side is not divergent as $z_4 \to z_3$, while we see from \eqref{ztwist} that $\lim_{z_4 \to z_3} |z_{34}|^{4/3} Z_{\text{twist}} = 1$, so that
\begin{equation}
    \lim_{z_4 \to z_3} |z_{34}|^{4/3} \Braket{\mathcal{O}_{\pi}(z_1) \mathcal{O}_{-\pi}(z_2) \mathcal{E}(z_3) \mathcal{E}(z_4)}_{\mathbb C} = |z_{12}|^{-4\lambda/5} = \Braket{\mathcal{O}_{\pi}(z_1) \mathcal{O}_{-\pi}(z_2)}_{\mathbb C}.
\end{equation}
Combining these observations gives the OPE
\begin{equation} \label{eq:OPE-ExE}
    {\mathcal E}(z) \times {\mathcal E}(z') = |z-z'|^{-4/3} \, \mathbb{1} + C^{\mathcal E}_{{\mathcal E}{\mathcal E}} \, |z-z'|^{-2/3} \, {\mathcal E}(z) + \ldots.
\end{equation}

Plugging this OPE into \eqref{eq:OOEE} and using \eqref{eq:OOE3} gives
\begin{align}
\begin{split}
    & \Braket{\mathcal{O}_\beta(z_1) \mathcal{O}_{-\beta}(z_2) \mathcal{E}(z_3) \mathcal{E}(z_4)}_{\mathbb C} \\
    & \quad = \Braket{\mathcal{O}_\beta(z_1) \mathcal{O}_{-\beta}(z_2)}_{\mathbb C} \, |z_{34}|^{-4/3} +  C^{\mathcal E}_{{\mathcal E}{\mathcal E}} \, \Braket{\mathcal{O}_\beta(z_1) \mathcal{O}_{-\beta}(z_2)\mathcal{E}(z_3)}_{\mathbb C} \, |z_{34}|^{-2/3} + O\big(|z_{34}|^{1/3}\big) \\
    & \quad = |z_{12}|^{-4\Delta(\beta)} |z_{34}|^{-4/3} + C^{\mathcal E}_{{\mathcal E}{\mathcal E}} \, C^{\mathcal E}_{{\mathcal O}_{\beta}{\mathcal O}_{-\beta}} \, |z_{12}|^{-4\Delta(\beta)} \, \left|\frac{z_{12}}{z_{13}z_{23}}\right|^{2/3} \, |z_{34}|^{-2/3} + O\big(|z_{34}|^{1/3}\big).
\end{split}
\end{align}
For $\beta=\pi$, comparing with \eqref{eq:OOEEpi} gives
\begin{align}
\begin{split} \label{eq:Ztwist+regular}
    & |z_{12}|^{-4\lambda/5} |z_{34}|^{-4/3} + C^{\mathcal E}_{{\mathcal E}{\mathcal E}} \, C^{\mathcal E}_{{\mathcal O}_{\pi}{\mathcal O}_{-\pi}} \, |z_{12}|^{-4\lambda/5} \, \left|\frac{z_{12}}{z_{13}z_{23}}\right|^{2/3} \, |z_{34}|^{-2/3} + O\big(|z_{34}|^{1/3}\big) \\
    & \quad = |z_{12}|^{-4\lambda/5} Z_{\text{twist}} + 4 \lambda \, |z_{12}|^{-4\lambda/5} \, \hat{\alpha}^{z_3}_{z_1|z_2;{\mathbb C}} \hat{\alpha}^{z_4}_{z_1|z_2;{\mathbb C}}.
\end{split}
\end{align}

Using the expression \eqref{ztwist} for $Z_{\text{twist}}$, we can write
\begin{align}
    & Z_{\text{twist}} = \left| \frac{z_{13} z_{24} }{ z_{23} z_{14} } \right|^{2/3} \left| {}_2 F_1\left( -\frac{2}{3}, \frac{1}{3}; \frac{2}{3}, x \right)  \right|^2 |z_{34}|^{-4/3} \nonumber \\
    & \quad - \left| \frac{ z_{12} }{ z_{23} z_{14} } \right|^{2/3} \frac{4\Gamma\left( \frac{2}{3} \right)^6}{\Gamma\left( \frac{4}{3} \right)^2 \Gamma\left( \frac{1}{3} \right)^4} \left| {}_2 F_1\left( -\frac{1}{3}, \frac{2}{3}; \frac{4}{3}, x \right) \right|^2 |z_{34}|^{-2/3}.
\end{align}
Plugging this into \eqref{eq:Ztwist+regular} and observing that
\begin{equation}
    \lim_{z_4 \to z_3} \left| \frac{z_{13} z_{24} }{ z_{23} z_{14} } \right|^{2/3} \left| {}_2 F_1\left( -\frac{2}{3}, \frac{1}{3}; \frac{2}{3}, x \right) \right|^2 = 1,
\end{equation}
shows that
\begin{equation}
    C^{\mathcal E}_{{\mathcal E}{\mathcal E}} \, C^{\mathcal E}_{{\mathcal O}_{\pi}{\mathcal O}_{-\pi}} = -\frac{4\Gamma\left( \frac{2}{3} \right)^6}{\Gamma\left( \frac{4}{3} \right)^2 \Gamma\left( \frac{1}{3} \right)^4} \lim_{z_3 \to z_4} \left| {}_2 F_1\left( -\frac{1}{3}, \frac{2}{3}; \frac{4}{3}, x \right) \right|^2 = -\frac{4\Gamma\left( \frac{2}{3} \right)^6}{\Gamma\left( \frac{4}{3} \right)^2 \Gamma\left( \frac{1}{3} \right)^4}.
\end{equation}
Finally, using \eqref{eq:C-EOO}, after some simplification we obtain 
\begin{align}
    C_{\mathcal{E}\mathcal{E}}^{\mathcal{E}} = \frac{1}{\sqrt{\lambda}} \frac{4 \cdot 2^{1/6} \cdot 3^{1/4} \cdot \sqrt{5} \pi^{3/2} \Gamma\left(\frac{2}{3}\right)}{\Gamma\left(\frac{1}{6}\right)^3 \Gamma\left(\frac{7}{6}\right)},
\end{align}
which indeed coincides with \eqref{CEEE}.

\section{Central charge} \label{sec:c}

Given an explicit form of a four-point function of a two dimensional CFT, together with sufficient knowledge of the operator spectrum, one can determine the central charge $c$ of the theory. We will now use the previous result \eqref{eq:OOEE} for the case of the full plane to confirm that $c=2 \lambda$ in the BLS, as was derived, for instance, in \cite{Camia_2016}.

In every two dimensional CFT, the two-point function of the energy–momentum tensor to leading order is fixed by conformal invariance to be
\begin{align} \label{tt1}
    \Braket{T(z_1) T(z_2)}_\mathbb{C} = \frac{c/2}{z_{12}^4}.
\end{align}
The energy-momentum tensor can be understood as the level-2 Virasoro descendant of the identity operator
\begin{align} \label{tt2}
    (L_{-2} \mathbb{1})(z) = \frac{1}{2\pi i} \oint_z d w \frac{1}{w-z} T(w) = T(z),
\end{align}
where the integral is along any contour around the point $z$, and $L_n$ are the generators of the Virasoro algebra. Its anti-holomorphic counterpart is analogously given by $\bar{T}(\bar{z}) = (\bar{L}_{-2} \mathbb{1})(\bar{z})$.

Additionally, the OPE of two primary operators is generally given by (cf.\ \cite{DiFrancesco:639405}, Section 6.6.3)
\begin{align}
\begin{split} \label{OPE}
    &\mathcal{A}_1(z+\epsilon) \times \mathcal{A}_2(z) \\
    =& \sum_\mathcal{P} \sum_{\{k,\bar{k}\}} C^\mathcal{P}_{12} \beta^{\mathcal{P}\{k\}}_{12} \bar{\beta}^{\mathcal{P}\{\bar{k}\}}_{12} \epsilon^{\Delta_\mathcal{P} - \Delta_1 - \Delta_2 + K} \bar{\epsilon}^{\bar{\Delta}_\mathcal{P} - \bar{\Delta}_1 - \bar{\Delta}_2 + \bar{K}} L_{-k_1}\ldots L_{-k_N} \bar{L}_{-\bar{k}_1}\ldots \bar{L}_{-\bar{k}_{\bar{N}}} \mathcal{P}(z),
\end{split}
\end{align}
where $C_{lj}^\mathcal{P}$ are three-point function coefficients, $K=\sum_{k_j \in \{k\}} k_j$ with $k_j \in \mathbb{N}$ is the descendant level, and $\beta_{l j}^{\mathcal{P}\{k\}}, \bar{\beta}_{l j}^{\mathcal{P}\{\bar{k}\}}$ are numerical coefficients that depend on the central charge and the conformal dimensions of the involved operators and are fully determined by the Virasoro algebra. The outer sum runs over all primary operators $\mathcal{P}$, and the inner sum is over all subsets $\{k\},\{\bar k\}$ of the natural numbers.
(This was  the basis of the analysis of Section \ref{sec:spectrum}.)

Since the identity operator has non-zero OPE coefficient for both $\mathcal{O}_\beta \times \mathcal{O}_{-\beta}$ and $\mathcal{E} \times \mathcal{E}$, we can use \eqref{eq:OOEE} to obtain the central charge $c$ by identifying the level-2 descendant of the identity.

We achieve this by applying the OPE twice to \eqref{eq:OOEE} and evaluating it in two equivalent ways.
First, we expand the expression
\begin{align} \label{TTexpansion1}
    \Braket{\mathcal{O}_\beta(z+\epsilon) \mathcal{O}_{-\beta}(z) \mathcal{E}(z'+\epsilon') \mathcal{E}(z')}_\mathbb{C}
\end{align}
analytically around zero for $\epsilon,\bar{\epsilon},\epsilon',\bar{\epsilon}'$.
We then identify the term of order $(\epsilon \epsilon')^{-\Delta(\beta) - 1/3 + 2}$ with the contribution from the algebraic expansion \eqref{OPE} at the same order in $\epsilon,\epsilon'$, which is
\begin{align} \label{TTexpansion2}
    (\epsilon \epsilon')^{-\Delta(\beta) - 1/3 + 2} C_{\mathcal{O}_\beta \mathcal{O}_{-\beta}}^\mathbb{1} C_{\mathcal{E} \mathcal{E}}^\mathbb{1}
    \beta_{\mathcal{O}_\beta \mathcal{O}_{-\beta}}^{\mathbb{1}\{2\}}
    \beta_{\mathcal{E} \mathcal{E}}^{\mathbb{1}\{2\}}
    \Braket{(L_{-2} \mathbb{1})(z) (L_{-2} \mathbb{1})(z')}_\mathbb{C}.
\end{align}
Generically, one expects contributions like $(L_{-1} \mathcal{A}^{(3,0)}) (L_{-1} \mathcal{A}^{(3,0)})$ and $\mathcal{A}^{(6,0)} \mathcal{A}^{(6,0)}$ to appear, where $\mathcal{A}^{(p,p')}$ are primary operators of conformal dimensions $(p/3,p'/3)$. However, the previous analysis has shown their relevant three-point coefficients vanish (see e.g.\ Figure \ref{VVEE_blocks}).

If the conformal dimensions of a pair of operators are equal, it can be shown that $\beta_{\mathcal{A}_1 \mathcal{A}_2}^{\mathbb{1}\{2\}} = 2 \Delta_{\mathcal{A}_1} / c$,
where $\Delta_{\mathcal{A}_1}=\Delta_{\mathcal{A}_2}$ is the conformal dimension of the operators \cite{DiFrancesco:639405}.
We also note that $C_{\mathcal{A}_1 \mathcal{A}_2}^\mathbb{1}$ denotes the normalization of non-zero two-point functions, which is canonically chosen to be 1. Every quantity in \eqref{TTexpansion2} has thus been determined.

The analytic expansion of \eqref{TTexpansion1} yields (at the desired order)
\begin{align} \label{cen1}
    (\epsilon \epsilon')^{-\Delta(\beta) - 1/3 + 2} \frac{1}{30} \frac{1-\cos\beta}{(z-z')^4}.
\end{align}
Using \eqref{tt1} and \eqref{tt2}, \eqref{TTexpansion2} becomes (dropping the powers of $\epsilon$ and $\epsilon'$)
\begin{align} \label{cen2}
    &\frac{2\Delta(\beta)}{c} \frac{2\Delta_\mathcal{E}}{c} \Braket{T(z)T(z')} = \frac{2}{3 c} \frac{\lambda}{10} \frac{1-\cos\beta}{(z-z')^4},
\end{align}
where we used $\Delta(\beta) = \frac{\lambda}{10}(1-\cos\beta), \Delta_\mathcal{E}=1/3$. Comparing \eqref{cen1} to \eqref{cen2} confirms the result that the BLS with intensity $\lambda$ has central charge $c = 2\lambda$.

\section{Conclusions and future work}

In this work we identified all scalar operators that couple to the layering vertex operators $\mathcal O_\beta$.  This leaves open the question of the nature of the operators with non-zero spin.  Perhaps the most interesting is the operator with $k=3, k'=0$ and zero charge, which has dimensions $(1,0)$.  This is a (component of a) spin-1 current that should satisfy a conservation law and generate a conserved charge.  
Understanding the nature and role of this current may greatly clarify the structure of the spectrum of the CFT associated to the BLS.

Another question open to investigation is the torus partition function.  By further exploiting the connection to the $O(n)$ model it seems possible that this can be computed.  If so it would reveal the complete spectrum and degeneracies of the theory (modulo complications resulting from the lack of unitarity of the theory).

The theory as we have presented it has a continuous spectrum because the operator dimensions depend on the continuous parameters $\beta$.  This is reminiscent of the vertex operators of the free boson.  There, one can compactify the boson and obtain a discrete spectrum.  An analogous procedure seems available here too, where we identify the layering number with itself modulo an integer.  If this is indeed self-consistent it would render the spectrum discrete, which has a number of interesting implications that we intend to explore in future work.

The  largest question is what place this Brownian loop soup conformal field theory has in the  spectrum of previously known conformally invariant models.  It appears to be a novel, self-consistent, and  rich theory in its own right, but its connections with the free field and the $O(n)$ model suggest that it may have ties to other theories that could be exploited to greatly advance our understanding of it.

\acknowledgments 

We are grateful to Sylvain Ribault for insightful comments on a draft of the manuscript.  The work of M.K.
is partially supported by the NSF through the grant PHY-1820814.

\appendix

\section{Proofs} \label{Appendix}

In this section we collect all the proofs that do not appear in the main body of the paper. We first show that the correlations functions $\langle E_{\varepsilon}(z_1) \ldots E_{\varepsilon}(z_n) \rangle_D$ are well defined, a necessary step to state Lemma \ref{lemma:n-point-function}, proved next in this appendix, and Theorem \ref{thm:edge}.
We then provide a proof of Lemma \ref{lemma:conf-cov-mu}.
We refer to Section \ref{sec:edge} for the notation used here, the statements of Lemmas \ref{lemma:n-point-function} and \ref{lemma:conf-cov-mu}, as well as the statement and proof of Theorem \ref{thm:edge}. Additionally, we remind the reader of the following definitions from Section \ref{sec:starmeasure}.

For any $\delta>0$, let ${\mathcal L}^{\delta}$ denote a Brownian loop soup in $D$ with intensity $\lambda$ and cutoff $\delta>0$, obtained by taking the usual Brownian loop soup and removing all loops with diameter smaller than $\delta$. We define $N^{\delta}_{\varepsilon}(z) \equiv n^{\varepsilon}_z({\mathcal L}^{\delta})$ and $E^{\delta}_{\varepsilon}(z) \equiv N^{\delta}_{\varepsilon}(z) - \langle N^{\delta}_{\varepsilon}(z) \rangle_D$.
Note that the random variables $N^{\delta}_{\varepsilon}(z)$ and $E^{\delta}_{\varepsilon}(z)$ are well defined because of the cutoffs $\varepsilon>0$ and $\delta>0$. The next lemma shows that, if we consider $n$-point functions of $E^{\delta}_{\varepsilon}$ for $n \geq 2$, the $\delta$ cutoff can be removed without the need to renormalize the $n$-point functions.
\begin{lemma} \label{lemma:cutoff-n-point-functions}
For any collection of points $z_1,\ldots,z_n \in D$ at distance greater than $2\varepsilon$ from each other, with $n \geq 2$, the following limit exists:
\begin{equation}
    \langle E_{\varepsilon}(z_1) \ldots E_{\varepsilon}(z_n) \rangle_D := \lim_{\delta \to 0} \langle E^{\delta}_{\varepsilon}(z_1) \ldots E^{\delta}_{\varepsilon}(z_n) \rangle_D .
\end{equation}
\end{lemma}

\noindent{\bf Proof.} For each $j=1,\ldots,n$, we can write
\begin{equation}
N^{\delta}_{\varepsilon}(z_j) = M^{\delta}_{\varepsilon}(z_j) + R^{\delta}_{\varepsilon}(z_j),
\end{equation}
where 
\begin{eqnarray}
    M^{\delta}_{\varepsilon}(z_j) & := & \sum_{\ell \in {\mathcal L}^{\delta}} \mathbf{I}(\ell \cap B_{\varepsilon}(z_j) \neq \emptyset, \ell \cap B_{\varepsilon}(z_k) = \emptyset \; \forall k \neq j), \\
    R^{\delta}_{\varepsilon}(z_j) & := & \sum_{\ell \in {\mathcal L}^{\delta}} \mathbf{I}(\ell \cap B_{\varepsilon}(z_j) \neq \emptyset \text{ and } \ell \cap B_{\varepsilon}(z_k) \neq \emptyset \text{ for at least one } k \neq j),
\end{eqnarray}
where $\mathbf{I}(\cdot)$ denotes the indicator function.

Now consider values of $\delta < \min_{k,m}(|z_k-z_m|-2\varepsilon)$ with $k,m=1,\ldots,n$ and $m \neq k$, then all the loops from $\mathcal L$ that intersect $B_{\varepsilon}(z_j)$ and at least one other disk $B_{\varepsilon}(z_k)$ must have diameter larger than $\delta$. Therefore, for $\delta$ sufficiently small, $R^{\delta}_{\varepsilon}(z_j)$ does not depend on $\delta$ and we can drop the superscript and write $R_{\varepsilon}(z_j)$ instead.

Defining $m^{\delta}_{\varepsilon}(z_j) := \langle M^{\delta}_{\varepsilon}(z_j) \rangle_D$ and $r_{\varepsilon}(z_j) := \langle R_{\varepsilon}(z_j) \rangle_D$, for values of $\delta$ sufficiently small we can write
\begin{align}
\begin{split}
    \Braket{E^{\delta}_{\varepsilon}(z_1) \ldots E^{\delta}_{\varepsilon}(z_n)}_D & = \Braket{ \left[M^{\delta}_{\varepsilon}(z_1) - m^{\delta}_{\varepsilon}(z_1) + R_{\varepsilon}(z_1) - r_{\varepsilon}(z_1) \right] E^{\delta}_{\varepsilon}(z_2) \ldots E^{\delta}_{\varepsilon}(z_n) }_D \\
    & = \Braket{ \left[M^{\delta}_{\varepsilon}(z_1) - m^{\delta}_{\varepsilon}(z_1) \right] E^{\delta}_{\varepsilon}(z_2) \ldots E^{\delta}_{\varepsilon}(z_n) }_D \\
    & \quad + \Braket{ \left[R_{\varepsilon}(z_1) - r_{\varepsilon}(z_1) \right] E^{\delta}_{\varepsilon}(z_2) \ldots E^{\delta}_{\varepsilon}(z_n) }_D.
\end{split}
\end{align}
$M^{\delta}_{\varepsilon}(z_1)$ is independent of $E^{\delta}_{\varepsilon}(z_j)$ for all $j \neq 1$, so
\begin{align}
    \Braket{ \left[M^{\delta}_{\varepsilon}(z_1) - m^{\delta}_{\varepsilon}(z_1)\right] E^{\delta}_{\varepsilon}(z_2) \ldots E^{\delta}_{\varepsilon}(z_n) }_D =0
\end{align}
and
\begin{align}
    \Braket{ E^{\delta}_{\varepsilon}(z_1) \ldots E^{\delta}_{\varepsilon}(z_n) }_D = \Braket{ \left[R_{\varepsilon}(z_1) - r_{\varepsilon}(z_1)\right] E^{\delta}_{\varepsilon}(z_2) \ldots E^{\delta}_{\varepsilon}(z_n) }_D.
\end{align}
Proceeding in the same way for all values of $j=2,\ldots,n$, we obtain
\begin{align}
    \Braket{ E^{\delta}_{\varepsilon}(z_1) \ldots E^{\delta}_{\varepsilon}(z_n) }_D = \Braket{ \left[R_{\varepsilon}(z_1) - r_{\varepsilon}(z_1)\right] \ldots \left[R_{\varepsilon}(z_n) - r_{\varepsilon}(z_n)\right] }_D,
\end{align}
which is independent of $\delta$. \qed

\medskip

\noindent{\bf Proof of Lemma \ref{lemma:n-point-function}.}
The random variables $(N^{\delta}_{\varepsilon}(z_1), \ldots, N^{\delta}_{\varepsilon}(z_n))$ are jointly Poisson.
If we let $\mathbf{v}=(v_1,\ldots,v_n)$ be an $n$-dimensional vector with components $v_j=0$ or $1$, following \cite{10.2996/kmj/1138036064} we see that their joint distribution is captured by
\begin{align}
\begin{split}
    N^{\delta}_{\varepsilon}(\mathbf{v}) := | \{ \ell: \text{diam}(\ell)>\delta, \ell \cap B_{\varepsilon}(z_j) \neq \emptyset \, \forall j: v_j=1, \ell \cap B_{\varepsilon}(z_j) = \emptyset \, \forall j: v_j=0 \} |,
\end{split}
\end{align}
where $N^{\delta}_{\varepsilon}(\mathbf{v})$ is itself a Poisson random variable with parameter $\lambda\mu^{\text{loop}}(\text{diam}(\ell)>\delta, \ell \cap B_{\varepsilon}(z_j) \neq \emptyset \, \forall j: v_j=1, \ell \cap B_{\varepsilon}(z_j) = \emptyset \, \forall j: v_j=0)$.
More precisely, using Theorem 2 of \cite{10.2996/kmj/1138036064}, we can write the joint probability generating function of $(N^{\delta}_{\varepsilon}(z_1), \ldots, N^{\delta}_{\varepsilon}(z_n))$ as
\begin{align}
\begin{split}
    &h(x_1,\ldots,x_n) := \Braket{x_1^{N^{\delta}_{\varepsilon}(z_1)}, \ldots, x_n^{N^{\delta}_{\varepsilon}(z_n)}} \\
    &\quad = \exp{\bigg[\lambda\sum_{\stackrel{I \text{ subset } \{1,\ldots,n\}}{|I| \geq 1}}} \mu^{\text{loop}}\big(\text{diam}(\ell)>\delta, \ell \cap B_{\varepsilon}(z_j) \neq \emptyset \, \forall j \in I, \ell \cap B_{\varepsilon}(z_j) = \emptyset \, \forall j \notin I \big) \\
    & \qquad \qquad \cdot \left( \prod_{j \in I} x_j - 1 \right) \bigg].
\end{split}
\end{align}
Letting $\mathcal{D}_k := \frac{\partial}{\partial x_k} - \lambda\mu^{\text{loop}}(\text{diam}(\ell)>\delta, \ell \cap B_{\varepsilon}(z_k) \neq \emptyset)$, using (\ref{E}) we have
\begin{align} \label{eq:differential-operator}
    \Braket{E^{\delta}_{\varepsilon}(z_1) \ldots E^{\delta}_{\varepsilon}(z_n)}_D = \left. \prod_{k=1}^n \mathcal{D}_k \, h(x_1,\ldots,x_n) \right|_{x_k=1}.
\end{align}

Using an induction argument, one can show that
\begin{align}
\begin{split}
    \sum_{\stackrel{I \text{ subset } \{1,\ldots,n\}}{|I| \geq 1}} \mu^{\text{loop}}\big(\text{diam}(\ell)>\delta, \ell \cap B_{\varepsilon}(z_j) \neq \emptyset \, \forall j \in I, \ell \cap B_{\varepsilon}(z_j) = \emptyset \, \forall j \notin I \big) \left( \prod_{j \in I} x_j - 1 \right) \\
    = \sum_{\stackrel{I \text{ subset } \{1,\ldots,n\}}{|I| \geq 1}} \mu^{\text{loop}}\big(\text{diam}(\ell)>\delta, \ell \cap B_{\varepsilon}(z_j) \neq \emptyset \, \forall j \in I \big) \prod_{j \in I}(x_j-1).
\end{split}
\end{align}
Hence,
\begin{align} \label{eq:prob-gen-func}
\begin{split}
    & h(x_1,\ldots,x_n) = \exp{\Big[\lambda\sum_{\stackrel{I \text{ subset } \{1,\ldots,n\}}{|I| \geq 1}} \mu^{\text{loop}}\big(\text{diam}(\ell)>\delta, \ell \cap B_{\varepsilon}(z_j) \neq \emptyset \, \forall j \in I \big) \prod_{j \in I}(x_j-1) \Big]} \\
    & = 1 + \sum_{r=1}^{\infty} \lambda^r \sum_{\stackrel{{I_1,\ldots,I_r}}{\text{subsets of }\{1,\ldots,n\}}} \prod_{l=1}^r \frac{1}{m(I_l)!} \Big(\mu^{\text{loop}}\big(\text{diam}(\ell)>\delta, \ell \cap B_{\varepsilon}(z_j) \neq \emptyset \, \forall j \in I_l \big) \prod_{j \in I_l}(x_j-1)\Big),
\end{split}
\end{align}
where the second sum is over all \emph{unordered} collections of subsets of $\{1,\ldots,n\}$ not necessarily distinct (i.e., over multiset), and we have used the fact that the number of ways in which an unordered collection of $r$ elements can be ordered is
\begin{align}
    \frac{r!}{\prod_{l=1}^r m(I_l)!},
\end{align}
where $m(I_l)$ is the multiplicity of $I_l$ in the multiset.

Considering the structure of the last expression, the definition of the differential operator $\mathcal{D}_k$, and the fact that in \eqref{eq:differential-operator}  all derivatives $\frac{\partial}{\partial x_k}$ are evaluated at $x_k=1$,  we can differentiate term by term.
It is clear that in the right-hand side of \eqref{eq:differential-operator} the only terms that survive are those for which the derivatives saturate the variables $x_k$. Moreover, Lemma \ref{lemma:cutoff-n-point-functions} implies that terms of the type $\mu^{\text{loop}}(\text{diam}(\ell)>\delta, \ell \cap B_{\varepsilon}(z_k))$ cannot be present in the right-hand side of \eqref{eq:differential-operator} because otherwise the limit $\delta \to 0$ would not exist. (One can reach the same conclusion by looking at \eqref{eq:prob-gen-func} and observing that terms containing subsets that are single points, i.e. $I_l=\{z_k\}$, disappear when applying $\mathcal{D}_k$.)
These considerations single out 
all partitions $\Pi$ of $\{1,\ldots,n\}$ whose elements have cardinality at least $2$.

Therefore, we obtain
\begin{align}
\begin{split}
    \Braket{E_{\varepsilon}(z_1) \ldots E_{\varepsilon}(z_n)}_D &= \left. \lim_{\delta \to 0}
    \prod_{k=1}^n \mathcal{D}_k h(x_1,\ldots,x_n) \right|_{x_k \equiv 1} \\
    &= \quad \sum_{\{I_1,\ldots,I_r\} \in \Pi} \lambda^r \prod_{l=1}^r \mu^{\text{loop}}_D(\ell \cap B_{\varepsilon}(z_j) \neq \emptyset \; \forall j \in I_l),
\end{split}
\end{align}
which concludes the proof. \qed

\medskip

\noindent{\bf Proof of Lemma \ref{lemma:conf-cov-mu}.}
Consider the full scaling limit of critical percolation in $D$ constructed in \cite{Camia_2006} and denote it by ${\mathcal F}_D$. ${\mathcal F}_D$ is distributed like CLE$_6$ in $D$ \cite{ea1b72ef0dc1467f9367b00129d7bf27}. As explained in \cite{2005math.....11605W}, the ``outer perimeters'' of loops from ${\mathcal F}_D$ are distributed like the outer boundaries of Brownian loops. Hence, there is a close connection between the Brownian loop measure $\mu^{\text{loop}}_D$ and the collection of loops constructed in \cite{Camia_2006}.

More precisely, let ${\mathbb P}$ denote the distribution of ${\mathcal F}_D$ and ${\mathbb E}$ denote expectation with respect to ${\mathbb P}$. Since ${\mathcal F}_D$ is conformally invariant, if $A$ is a measurable set of self-avoiding loops and ${\mathcal N}_{A}$ is the number of loops $\Gamma$ from ${\mathcal F}_D$ such that their outer perimeters $\ell(\Gamma)$ are in $A$, ${\mathbb E}({\mathcal N}_{A})$ defines a conformally invariant measure on self-avoiding loops. Moreover, since the measure $\mu^{\text{loop}}_D$ is unique, up to a multiplicative constant, we must have
\begin{equation} \label{eq:mu-perc}
    \mu^{\text{loop}}_D(A) = \Theta \, {\mathbb E}({\mathcal N}_{A}),
\end{equation}
where $0<\Theta<\infty$ is a constant.

Now consider the set of simple loops $S_{\varepsilon} = \{ \ell \in D: \ell \cap B_{\varepsilon}(z_j) \neq \emptyset \; \forall j=1,\ldots,k \}$. Thanks to the scale invariance of $\mu^{\text{loop}}_D$ and ${\mathcal F}_D$, we can assume without loss of generality that the points $z_1,\ldots,z_k$ are at distance much larger than $1$ from each other and from $\partial D$. We write $\mathcal{F}_D \in S_{\varepsilon}$ to indicate the event that a configuration from $\mathcal{F}_D$ contains at least one loop $\Gamma$ such that $\ell(\Gamma) \in S_{\varepsilon}$.

For each $j=1,\ldots,k$, consider the annulus $A_{\varepsilon,1}(z_j) := B_{1}(z_j) \setminus B_{\varepsilon}(z_j)$ centered at $z_j$ with outer radius $1$ and inner radius $\varepsilon$. Because of our assumption on the distances between the points $z_j, j=1,\ldots,k$, the annuli do not overlap. The configurations from ${\mathcal F}_D$ for which $\mathcal{N}_{S_{\varepsilon}} > 0$ (i.e., such that $\mathcal{F}_D \in S_{\varepsilon}$) are those that contain at least one loop $\Gamma$ whose outer perimeter $\ell(\Gamma)$ intersects $B_{\varepsilon}(z_j)$ for each $j=1,\ldots,k$. They can be split in two groups as described below, where a \emph{three-arm event} inside $A_{\varepsilon,1}(z_j)$ refers to the presence of a loop $\Gamma$ such that the annulus $A_{\varepsilon,1}(z_j)$ is crossed from the inside of $B_{\varepsilon}(z_j)$ to the outside of $B_{1}(z_j)$ by two disjoint outer perimeter paths belonging to $\ell(\Gamma)$ and by one path within the complement of the unique unbounded component of $\mathbb{C}\setminus\Gamma$.
\begin{enumerate}
    \item[(i)] Configurations that induce a three-arm event inside $A_{\varepsilon,1}(z_j)$ for each $j=1,\ldots,k$, for which ${\mathcal N}_{S_{\varepsilon}}=1$.
    \item[(ii)] Configuration that induce more than three arms in $A_{\varepsilon,1}(z_j)$ for at least one $j=1,\ldots,k$, for which ${\mathcal N}_{S_{\varepsilon}} \geq 1$.
\end{enumerate}

The probability of a three-arm event in $A_{\varepsilon,1}(z_j)$ is $\vartheta_{\varepsilon}\sim\varepsilon^{2/3}$, while the probability to have four or more arms in $A_{\varepsilon,1}(z_j)$ is $o(\vartheta_{\varepsilon})$ as $\varepsilon \to 0$; therefore
\begin{eqnarray}
     \vartheta_{\varepsilon}^{-k} {\mathbb E}({\mathcal N}_{S_{\varepsilon}}) & = & \vartheta_{\varepsilon}^{-k} {\mathbb P}(\mathcal{F}_D \in S_{\varepsilon} \text{ and there is a three-arm event in each } A_{\varepsilon,1}(z_j)) \nonumber \\
     & & + \, o(\varepsilon). \label{O(7/12)}
\end{eqnarray}

It follows from the construction of ${\mathcal F}_D$ in \cite{Camia_2006}, which uses the locality of SLE$_6$, that a configuration in group (i) can be constructed by first generating independent configurations inside $B_{1}(z_j)$ for each $j=1,\ldots,k$, requiring that each induces a three-arm event in $A_{\varepsilon,1}(z_j)$, and then generating a ``matching'' configuration in $D \setminus \cup_{j=1}^k B_{1}(z_j)$. A configuration inside $B_{1}(z_j)$ contains loops and arcs starting and ending on $\partial B_{1}(z_j)$. Moreover, since $A_{\varepsilon,1}(z_j)$ contains a three-arm event, exactly one outer perimeter arc starting and ending on $\partial B_{1}(z_j)$ intersects $B_{\varepsilon}(z_j)$. Each arc in $B_{1}(z_j)$ has a pair of endpoints on $\partial B_{1}(z_j)$. We let $\mathcal{I}_j$ denote the collection of endpoints on $\partial B_{1}(z_j)$, together with the information regarding which endpoints are connected to each other, and we denote by $\nu_j^{\varepsilon}$ the distribution of $\mathcal{I}_j$, conditioned on the occurrence of a three-arm event in $A_{\varepsilon,1}(z_j)$.
An important observation is that, conditioned on $\mathcal{I}_j$ for each $j=1,\ldots,k$, the configuration in $D \setminus \cup_{j=1}^k B_{1}(z_j)$ is independent of the configurations inside $B_{1}(z_j)$ for $j=1,\ldots,k$. If we let $G$ denote the event that endpoints on $\partial B_{1}(z_j)$ are connected in $D \setminus \cup_{j=1}^k B_{1}(z_j)$ in such a way that overall the resulting configuration in $D$ is in $ S_{\varepsilon}$, this observation allows us to write
\begin{align}
\begin{split}
     {\mathbb P}(\mathcal{F}_D \in S_{\varepsilon} \text{ and there is a three-arm event in } A_{\varepsilon,1}(z_j) \, \forall j=1,\ldots,k)& \\
     = {\mathbb P}(\mathcal{F}_D \in S_{\varepsilon} | \text{ there is a three-arm event in } A_{\varepsilon,1}(z_j) \, \forall j=1,\ldots,k)& \\
     {\mathbb P}(\text{there is a three-arm event in } A_{\varepsilon,1}(z_j) \, \forall j=1,\ldots,k)& \\
     = \vartheta_{\varepsilon}^{k} \int {\mathbb P}(G | \mathcal{I}_1,\ldots,\mathcal{I}_k) \prod_{j=1}^k d\nu^{\varepsilon}_j(\mathcal{I}_j)&.
\end{split}
\end{align}

Combining this with \eqref{O(7/12)}, we obtain
\begin{align} \label{eq:integral}
     \lim_{\varepsilon \to 0} \vartheta_{\varepsilon}^{-k} {\mathbb E}({\mathcal N}_{S_{\varepsilon}}) = \lim_{\varepsilon \to 0} \int {\mathbb P}(G | \mathcal{I}_1,\ldots,\mathcal{I}_k) \prod_{j=1}^k d\nu^{\varepsilon}_j(\mathcal{I}_j),
\end{align}
where ${\mathbb P}(G \vert \mathcal{I}_1,\ldots,\mathcal{I}_k)$ does not depend on $\varepsilon$ and $\nu^{\varepsilon}_j$ is the distribution of endpoints on $\partial B_{1}(z_j)$ conditioned on the occurrence of a three-arm event in $A_{\varepsilon,1}(z_j)$, or equivalently on the existence of a single outer perimeter arc starting and ending on $\partial B_1(z_j)$ and intersecting $B_{\varepsilon}(z_j)$.

Now observe that requiring the existence of a single outer perimeter arc that intersects $B_{\varepsilon}(z_j)$ and sending $\varepsilon \to 0$ is equivalent to centering the disk $B_{1}(z_j)$ at a \emph{typical} point\footnote{Here typical means that it is not a \emph{pivotal} point, i.e., a point on the outer perimeter of two loops. Pivotal points have a lower fractal dimension.} $z_j$ on the outer perimeter of a loop from ${\mathcal F}_D$ which exits $B_{1}(z_j)$ and therefore has diameter greater than $1$. Therefore, the limit $\lim_{\varepsilon \to 0} \nu^{\varepsilon}_j$ exists: it is given by the distribution of endpoints of arcs for a disk of radius $1$ centered at a typical point on the outer perimeter of a loop from ${\mathcal F}_D$ of diameter larger than $1$. Equivalently, by scale invariance, it is the distribution of endpoints of arcs on $\partial B_r(z)$ for a disk $B_r(z)$ centered at a typical point $z$ on the outer perimeter of a loop from ${\mathcal F}_D$, with diameter $r$ smaller than the diameter of the loop. Therefore, if we call $\nu$ this distribution, from \eqref{eq:mu-perc} and \eqref{eq:integral} we have
\begin{align}
\begin{split}
    \lim_{\varepsilon \to 0} \vartheta_{\varepsilon}^{-k} \mu^{\text{loop}}_D(S_{\varepsilon}) & = \Theta \lim_{\varepsilon \to 0} \vartheta_{\varepsilon}^{-k} {\mathbb E}({\mathcal N}_{S_{\varepsilon}}) \\
    & = \Theta \int {\mathbb P}(G | \mathcal{I}_1,\ldots,\mathcal{I}_k) \prod_{j=1}^k d\nu(\mathcal{I}_j),
\end{split}
\end{align}
proving the existence of the limit in \eqref{eq:epsilon-limit}.

In order to prove \eqref{eq:conf-cov-mu}, consider a domain $D'$ conformally equivalent to $D$ and a conformal map $f:D \to D'$, and let $z'_j=f(z_j), s_j=|f'(z_j)|$ for each $j=1,\ldots,k$, and $S'_{\varepsilon} = \{ \ell \in D': \ell \cap B_{\varepsilon}(z'_j) \neq \emptyset \; \forall j=1,\ldots,k \}$. We are interested in the behavior of
\begin{equation} \label{eq:limit}
    \alpha^{z'_1,\ldots,z'_k}_{D'} = \lim_{\varepsilon \to 0} \vartheta_{\varepsilon}^{-k} \mu_{D'}(S'_{\varepsilon}) = \lim_{\varepsilon \to 0} \vartheta_{\varepsilon}^{-k} \mu_D(\ell \cap f^{-1}(B_{\varepsilon}(z'_j)) \neq \emptyset \; \forall j=1,\ldots,k).
\end{equation}
To evaluate this limit, we will use the fact that
\begin{align}
\begin{split} \label{eq:mu-mu}
    &\vartheta_{\varepsilon}^{-k} \, \mu_D(\ell \cap f^{-1}(B_{\varepsilon}(z'_j)) \neq \emptyset \; \forall j=1,\ldots,k) - \mu_D(\ell \cap B_{\varepsilon/s_j}(z_j) \neq \emptyset \; \forall j=1,\ldots,k) \\
    & \qquad = o(1) \text{ as } \varepsilon \to 0.
\end{split}
\end{align}
To see this, let $A_{r_j,R_j}(z_j) = B_{R_j}(z_j) \setminus B_{r_j}(z_j)$ denote the thinnest annulus centered at $z_j$ containing the symmetric difference of $f^{-1}(B_{\varepsilon}(z'_j))$ and $B_{\varepsilon/s_j}(z_j)$ and note that
\begin{align}
\begin{split} \label{eq:upper-bound}
    &\big\vert \mu_D(\ell \cap f^{-1}(B_{\varepsilon}(z'_j)) \neq \emptyset \; \forall j=1,\ldots,k) - \mu_D(\ell \cap B_{\varepsilon/s_j}(z_j) \neq \emptyset \; \forall j=1,\ldots,k) \big\vert \\
    & \quad \leq \mu_D(\ell \cap B_{R_j}(z_j) \neq \emptyset \;\forall j=1,\ldots,k \text{ and } \ell \cap B_{r_l}(z_l) = \emptyset \text{ for at least one } l=1,\ldots,k ).
\end{split}
\end{align}
Since $f^{-1}$ is analytic and $(f^{-1}(z'_j))'=1/s_j$, for every $w \in \partial B_{\varepsilon}(z'_j)$, $|z_j-f^{-1}(w)| = |f^{-1}(z'_j)-f^{-1}(w)| = \varepsilon/s_j + O(\varepsilon^2)$, which implies that $R_j-r_j=O(\varepsilon^2)$ and $R_j=O(\varepsilon)$.
The second line of \eqref{eq:upper-bound} can be bounded above by a constant times $\vartheta_{\varepsilon}^{k} \times o(1)$, as we now explain.
The factor $\vartheta_{\varepsilon}^{k}$ comes from the requirement that $\ell$ intersect
 $B_{R_{z_j}}(z_j)$ for each $j=1,\ldots,k$ and the factor $o(1)$ comes from the requirement that $\ell$ intersect $\partial B_{R_l}(z_l)$ but not $\partial B_{r_l}(z_l)$ for at least one $l=1,\ldots,k$. 
More precisely, one can consider disks $D_j$ of radius $N\varepsilon$ centered at $z_1,\ldots,z_k$, for some $N$ large but fixed, and first explore the region outside these disks. Using percolation arguments similar to those in the first part of the proof, one gets a factor $\vartheta_{N\varepsilon}^{k}=O\big(\vartheta_{\varepsilon}^{k}\big)$ from the requirement that $\ell$ intersect each $D_j$. Inside each disk $D_j$, one has a Brownian excursion of linear size $N\varepsilon$ that gets to distance $O(\varepsilon^2)$ of $\partial B_{r_j}(z_j)$ without intersecting it. The $\mu_D^{\text{loop}}$-measure of loops producing such excursions can be shown to be of order $o(1)$, as $\varepsilon \to 0$, by arguments similar to those in the proof of Lemma 6.5 of \cite{2015PTRF}, which provides an upper bound for the probability that a Brownian loop gets close to a deterministic loop without touching it. The upper bound implies that the probability in question goes to zero when the ratio between the linear size of the deterministic loop and the minimal distance between the loops diverges, provided that the Brownian loop has linear size comparable to that of the deterministic loop. In the present case, that ratio is of order $1/\varepsilon$ and the Brownian excursion has diameter of order $N\varepsilon$, comparable to the diameter of $\partial B_{r_j}(z_j)$.

Hence, from \eqref{eq:limit}, \eqref{eq:mu-mu} and \eqref{eq:epsilon-limit}, using Lemma \ref{lemma:theta} below, we obtain
\begin{align}
\begin{split}
    \alpha^{z'_1,\ldots,z'_k}_{D'} & = \lim_{\varepsilon \to 0} \big( \vartheta_{\varepsilon}^{-k} \mu_D(\ell \cap B_{\varepsilon/s_j}(z_j)) \neq \emptyset \; \forall j=1,\ldots,k) + o(1) \big) \\
    & = \left( \prod_{j=1}^k s_j^{-2/3} \right) \lim_{\varepsilon \to 0} \left( \prod_{j=1}^k \left(\frac{\vartheta_{\varepsilon}/\vartheta_{\varepsilon/s_j}}{s_j^{2/3}} \right)^{-1} \vartheta_{\varepsilon/s_j}^{-1} \right) \mu_D(\ell \cap B_{\varepsilon/s_j}(z_j)) \neq \emptyset \; \forall j=1,\ldots,k) \\
    & = \left( \prod_{j=1}^k s_j^{-2/3} \right) \alpha^{z_1,\ldots,z_k}_{D},
\end{split}
\end{align}
which concludes the proof, modulo the proof of Lemma \ref{lemma:theta}, provided below. \qed

\medskip

\begin{lemma} \label{lemma:theta}
For any $s>0$ we have
\begin{align} \label{eq:theta}
    \lim_{\varepsilon \to 0}\frac{\vartheta_{\varepsilon}}{\vartheta_{\varepsilon/s}} = s^{2/3}.
\end{align}
\end{lemma}

\noindent{\bf Proof.}
If \eqref{eq:theta} holds for $s<1$, for $s>1$, letting $r=1/s$, by scale invariance we have
\begin{align}
        \lim_{\varepsilon \to 0}\frac{\vartheta_{\varepsilon}}{\vartheta_{\varepsilon/s}} = \lim_{\varepsilon \to 0}\frac{\vartheta_{\varepsilon/r}}{\vartheta_{\varepsilon}} = r^{-2/3} = s^{2/3}.
\end{align}
Hence, it is enough to prove \eqref{eq:theta} for $s<1$ and so in the rest of the proof we assume that $s<1$.
We use the notation introduced in the proof of Lemma \ref{lemma:conf-cov-mu} and further let $\vartheta(\varepsilon,s)$ denote the probability of a three-arm event in an annulus with inner radius $\varepsilon$ and outer radius $s$. (In particular, $\vartheta(\varepsilon,1)\equiv\vartheta_{\varepsilon}$.) 
We will show that
\begin{align} \label{eq:theta-limit}
    \lim_{\varepsilon \to 0}\frac{\vartheta(\varepsilon,1)}{\vartheta(\varepsilon,s)} = s^{2/3}.
\end{align}
By scale invariance, this implies that
\begin{align}
    \lim_{\varepsilon \to 0}\frac{\vartheta(\varepsilon,1)}{\vartheta(\varepsilon/s,1)} = s^{2/3},
\end{align}
as desired.
For any $\varepsilon<s<1$, we will let $\vartheta(\varepsilon,1 \, \vert \, \varepsilon,s)$ denote the conditional probability of a three-arm event in $A_{\varepsilon,1}(0)$, given the existence of a three-arm event in $A_{\varepsilon,s}(0)$.
The existence of the limit in \eqref{eq:theta-limit} follows from the scale invariance of the scaling limit of percolation. Using the notation introduced in the proof of Lemma \ref{lemma:conf-cov-mu}, the scale invariance of the percolation scaling limit implies the scale invariance of $\nu^{\varepsilon}$, which allows us to write
\begin{align}
\begin{split} \label{eq:limit-exists}
        \lim_{\varepsilon \to 0}\frac{\vartheta(\varepsilon,1)}{\vartheta(\varepsilon,s)} &= \lim_{\varepsilon \to 0}\vartheta(\varepsilon,1 \, \vert \, \varepsilon,s) \\
        &= \lim_{\varepsilon \to 0}\int {\mathbb P}(H | \mathcal{I}) d\nu^{\varepsilon/s}(\mathcal{I}) \\
        &= \int {\mathbb P}(H | \mathcal{I}) d\nu(\mathcal{I}) =: L,
\end{split}
\end{align}
where $H$ is the event that a loop responsible for a three-arm event in $A_{\varepsilon/s,1}(0)$ reaches $\partial B_{1/s}(0)$, thus producing a three-arm event in $A_{\varepsilon/s,1/s}(0)$, which has the same probability as a three-arm event in $A_{\varepsilon,1}(0)$.
Now that we know that the limit exists, \eqref{eq:theta-limit} can be obtained as in the proof of the second limit in Equation (4.28) of Proposition 4.9 of \cite{garban2013pivotal}. We repeat the argument here for the reader's convenience.
It is known that $\vartheta_{\varepsilon}=\varepsilon^{2/3+o(1)}$, where $o(1)$ goes to zero as $\varepsilon \to 0$, so that
\begin{align} \label{eq:lim-log}
    \lim_{n \to \infty}\frac{\log\vartheta(s^n,1)}{n} = \log s^{2/3}.
\end{align}
Now note that $\vartheta(s^n,1)$ can be written as
\begin{align}
    \vartheta(s^n,1) = \frac{\vartheta(s^n,1)}{\vartheta(s^n,s)} \frac{\vartheta(s^{n-1},1)}{\vartheta(s^{n-1},s)} \ldots \frac{\vartheta(s,1)}{1},
\end{align}
which implies that
\begin{align} \label{eq:sum-logs}
        \frac{\log\vartheta(s^n,1)}{n} = \frac{1}{n} \sum_{j=1}^n \log\frac{\vartheta(s^j,1)}{\vartheta(s^j,s)}.
\end{align}
Since $s<1$, using \eqref{eq:limit-exists}, we have
\begin{align}
    \lim_{j \to \infty}\log\frac{\vartheta(s^j,1)}{\vartheta(s^j,s)} = \log L.
\end{align}
By convergence of the Ces\`aro mean, the right-hand side of \eqref{eq:sum-logs} converges to $\log L$, so \eqref{eq:sum-logs} and \eqref{eq:lim-log} imply $\log L = \log s^{2/3}$, which concludes the proof.

\qed

\medskip

\noindent{\bf Proof of Lemma \ref{lemma:correlations-higher-cutoff}}
This proof is similar to that of Lemma \ref{lemma:n-point-function}.
With the notation introduced in the proof of Lemma \ref{lemma:n-point-function}, we have that
\begin{align} \label{eq:differential-operator-higher}
    \Braket{E^{(k_1);\delta}_{\varepsilon}(z_1) \ldots E^{(k_n);\delta}_{\varepsilon}(z_n)}_D = \left. \prod_{j=1}^n \mathcal{D}_j^{k_j} \, h(x_1,\ldots,x_n) \right|_{x_j \equiv 1}.
\end{align}
Considering the structure of \eqref{eq:prob-gen-func}, the definition of the differential operator $\mathcal{D}_j$, and the fact that in \eqref{eq:differential-operator-higher} all derivatives $\frac{\partial}{\partial x_j}$ are evaluated at $x_j=1$, it is clear that in the right-hand side of \eqref{eq:differential-operator-higher} the only terms that survive are those for which the derivatives saturate the variables $x_j$. Moreover, the structure of \eqref{eq:prob-gen-func} implies that all terms containing subsets that are single points, i.e.\ $I_l=\{z_j\}$, disappear when applying $D_j$.
These considerations imply that the only non-zero terms are those corresponding to multisets $M \in \mathcal{M}$. Note also that, when $\frac{\partial}{\partial x_j}$ is applied $k_j$ times to $h(x_1,\ldots,x_n)$, as prescribed by $\mathcal{D}_j^{k_j}$ it produces a multiplicative factor $k_j!$ for each $j=1,\ldots,n$.

Therefore, if the vector $\mathbf{k}=(k_1, \ldots, k_n)$ is such that $\mathcal{M} = \emptyset$, we obtain
\begin{align}
\begin{split}
    & \Braket{E_{\varepsilon}(z_1) \ldots E_{\varepsilon}(z_n)}_D = \left. \lim_{\delta \to 0}
    \prod_{j=1}^n \mathcal{D}_j^{k_j} h(x_1,\ldots,x_n) \right|_{x_j \equiv 1} \\
    & = \Big( \prod_{j=1}^n k_j! \Big) \sum_{M \in \mathcal{M}} \lambda^{\sum_{S \in M} m_M(S)} \prod_{S \in M} \frac{1}{m_M(S)!} \Big(\mu^{\text{loop}}_D(\ell \cap B_{\varepsilon}(z_j) \neq \emptyset \; \forall j \in I_S)\Big)^{m_M(S)},
\end{split}
\end{align}
otherwise we get zero, as required. \qed

\bibliographystyle{unsrt}
\bibliography{bibliography}

\end{document}